\documentclass{cernrep}
\usepackage{color}
\usepackage{tabularx}
\usepackage{mathtools}
\usepackage[flushleft]{threeparttable}
\usepackage[unicode=true,pdfusetitle,
 bookmarks=true,bookmarksnumbered=false,bookmarksopen=false,
 breaklinks=false,pdfborder={0 0 1},backref=false,colorlinks=true,linkcolor=blue,linktoc=page]
 {hyperref}
\usepackage{authblk}
\usepackage{xspace,amsmath,amssymb}
\makeatletter
\usepackage{hyperref}
\newcommand\footnoteref[1]{\protected@xdef\@thefnmark{\ref{#1}}\@footnotemark}
\makeatother
\usepackage{ifthen} 
\newboolean{articletitles}
\setboolean{articletitles}{true} 
\usepackage{amsmath}

\newcommand{\GeV}{\ensuremath{\text{GeV}}\xspace}
\newcommand{\fb}{\ensuremath{\text{fb}}\xspace}

\newcommand{\mjj}{\ensuremath{m_{\mathrm{jj}}}\xspace}

\newcommand*{\SCIKITLEARN}[1]{\textsc{Scikit-learn}~#1\xspace}

\begin{document}
\title{Gaussian process regression as a sustainable data-driven background estimate method at the (HL)-LHC}

\author[1,2]{Jackson Barr}
\author[3]{Bingxuan Liu}
\affil[1]{Centre for Data Intensive Science and Industry, University College London}
\affil[2]{Deutsches Elektronen-Synchrotron DESY}
\affil[3]{School of Science, Shenzhen Campus of Sun Yat-sen University}

\begin{abstract}

In this article, we evaluate the performance of a data-driven background
estimate method based on Gaussian Process Regression (GPR). A realistic
background spectrum from a search conducted by CMS is considered, where a large
sub-region below the trigger threshold is included. It is found that the $L_2$
regularisation can serve as a set of hyperparameters and control the overall
modelling performance to satisfy common standards established by experiments at
the Large Hadron Collider (LHC). In addition, we show the robustness of this
method against increasing luminosity via pseudo-experiments matching the
expected luminosity at the High-Luminosity LHC (HL-LHC). While traditional
methods relying on empirical functions have been challenged during LHC Run 2
already, a GPR-based technique can offer a solution that is valid through the
entire lifetime of the (HL)-LHC. 

\end{abstract}


\maketitle
\tableofcontents
\clearpage
\section{{\bfseries Introduction}}
\label{sec:intro} 

Background modelling plays a pivotal role in searches for new
physics~\cite{ATLASExo,CMSdarksector,CMStlareview} and precision measurements
of the Standard Model~\cite{ATLASSM, ATLASTop}, carried out by the major
experiments at the Large Hadron Collider (LHC). Thanks to the careful tuning
and calibrations, simulated event samples can describe the data at the required
level of precision for most cases~\cite{powhegbox,pythia8,Sherpa}. Events
originated from Quantum Chromodynamics (QCD) processes have a large jet
multiplicity, and it is the dominating background contribution in numerous
analyses considering hadronic final states. The modelling of those multijet
events is known to be suboptimal. Due to its massive cross-section in
$pp$ collisions, the simulation is subject to significant statistical
uncertainties. In addition, the theoretical uncertainties are not sufficient to
ensure desired precision across the entire phase
space~\cite{powhegdijet,powhegtrijet,Buckley,dijethadronization,CMSjetmeasurement,ATLASjetxs,trijetlhc}.
To overcome this challenge, various experiments have developed data-driven
methods to estimate the multijet background in physics analyses. In the pursuit
of heavy particles with narrow widths, i.e., analyses looking for bumps, a
common background estimation strategy is to apply a functional fit to the data
spectrum. An empirical function can fit the background distribution,
while a narrow peak over the background cannot be incorporated into the
function. A widely used function has the following form:

\begin{equation}
f(x) = p_{0}(1-x)^{p_1}x^{p_2 + p_3\ln x + p_4(\ln x)^2 + ...}   
\label{equ:dijet}
\end{equation}

where $x$ is a scaled variable defined as $x = m /\sqrt{s}$.
$m$ is the mass observable, such as $\mjj$, the invariant mass of the
di-jet system. Variations of the above function are also viable
options~\cite{ATLASDijet,ATLASTLA,CMSTLA,ATLASdiphoton}. Depending on how large
and how complex the dataset is, one can decide how many higher order
logarithmic terms to include. Exponential functions and Bernstein polynomials
have been applied in analyses as well~\cite{UA2_1,UA2_2,CMSdiphoton}. 

This methodology has been quite successful, but the unprecedented integrated
luminosity recorded by the LHC starts challenging it. In fact, several recent
analyses reported that this function family cannot easily handle the
large datasets any more, so a sliding window technique is introduced, where
individual fit is performed for each bin using a subset of the
spectrum~\cite{ATLASDijet,ATLASTLA}. In addition to the increasing luminosity,
the expanding search programme also demands a more universal strategy not
relying on empirical functions. Naturally, the community has started
re-thinking about the functional fit approach and investigating completely
alternative methodologies.

In ref.~\cite{FD}, a method based on orthonormal series is constructed, and it
is successfully applied in an ATLAS analysis as the primary background
estimate~\cite{ATLASMultib}. It does not rely on empirical functional forms, and
it is mathematically sound. With a complete orthonormal basis, an arbitrary
spectrum can be described as long as there are enough terms. Though it is more
general compared to the canonical functional fit, the authors of ref.~\cite{FD}
had to come up with a new basis that is more suitable for HEP experiments. The
new method developed in ref.~\cite{SymbolFit} uses symbolic regression to
automate parametric modelling, which shows great flexibility as well. Methods in
ref.~\cite{FD} and ref.~\cite{SymbolFit} are both parametric in their final
applications.

Gaussian Process Regression (GPR), on the other hand, is a non-parametric
technique broadly used in machine learning and statistics~\cite{GPR2006}. A
Gaussian Process is a collection of random variables such that any finite
combination of them has a joint Gaussian distribution. Its potential usage in
HEP experiments is discussed in ref.~\cite{GPR2017,GPR2022} and several LHC
results have applied this method to achieve various goals, such as template
smoothing~\cite{ATLASyy} or background
estimation~\cite{CMSTLA,ATLASHC}. In a GPR considering binned data, the bin contents $y_1 ...
y_n$ are described by a Gaussian PDF:\[ p(y_i; \mu_i, \mathbf{K} ) =
\frac{1}{\sqrt{(2\pi)^n |\mathbf{K}|}} \exp \left( -\frac{1}{2}\sum_{i,j}^n (y_i
- \mu_i) K^{-1}_{ij} (y_j - \mu_j)  \right), \] where $\mu_i$ is the prior mean of $y_i$, and $\mathbf{K}$ is the covariance matrix,
parameterised by a kernel function $K(x_i, x_j)$. A common choice of kernel
is the Radial Basis Function (RBF) kernel:

\[ K(x_i,x_j) = \exp\left(- \frac{\lvert \lvert x_i - x_j \rvert
\rvert^2}{2\ell^2} \right), \] 

where $\ell$ is the length scale of the kernel and $x_{i(j)}$ refers to the bin
centre of the training data. It determines how closely correlated the neighbouring
points are and can be fit to data through maximising the marginal likelihood of
the observations. The fitted model gives us the posterior distribution of $y_i$, which can be used to calculate the mean and the standard deviation of the estimate. The authors of ref.~\cite{GPR2017} and ref.~\cite{GPR2022}
explored the application of GPR in di-jet resonance searches. While previous
work was concentrated on testing different kernels, in this work, more
attention is paid to the hyperparameters, the pre-defined parameters that are
not optimised by GPR. It is demonstrated that the $L_2$ regularisation, which
represents the level of noise in the input dataset, can be introduced as a set of
hyperparameters to control the overall performance of GPR. Applying the widely
adopted RBF kernel, with minimal hyperparameter tuning, GPR is capable of
fitting a complicated spectrum that is challenging for functional fit methods.
It is also robust against the increasing luminosity up to the HL-LHC era.

The article is organised as follows: Section~\ref{sec:dataset} details the test
datasets in this work; Section~\ref{sec:tuning} discusses the optimal
representation of the data and the GPR model setup;
Section~\ref{sec:sensitivity} lays out a comprehensive study of the expected
sensitivity, followed by a similar study in the context of the HL-LHC in
Section~\ref{sec:hllhc}; and finally, Section~\ref{sec:summary} gives a
summary. 

\section{{\bfseries Test dataset}}
\label{sec:dataset}

\begin{figure}[ht]
  \begin{center}
   \includegraphics[width=0.4\columnwidth]{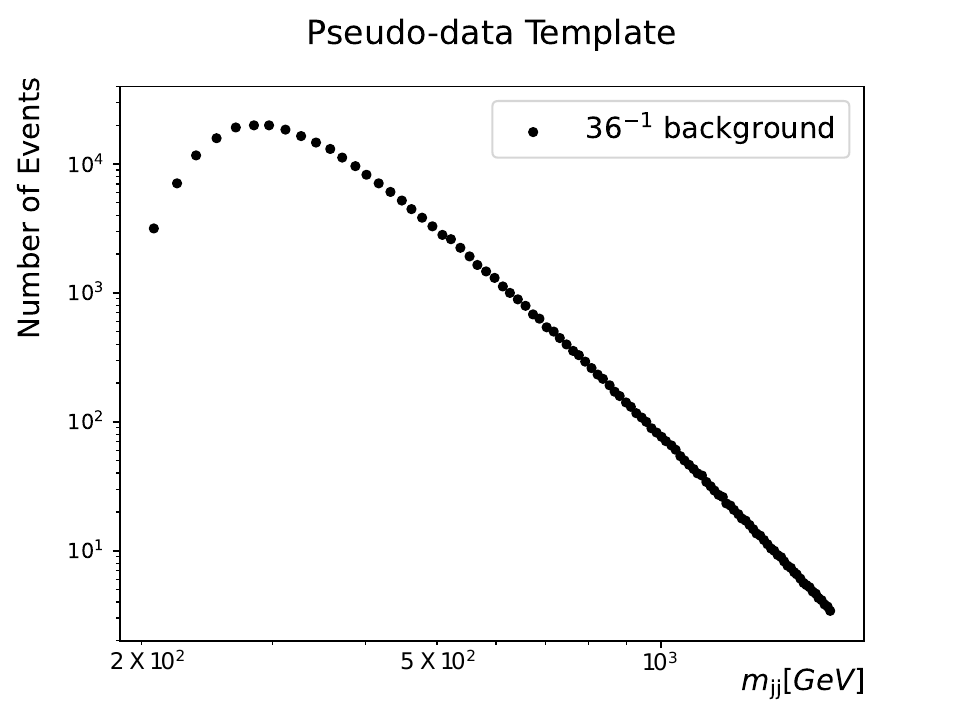}
   \includegraphics[width=0.4\columnwidth]{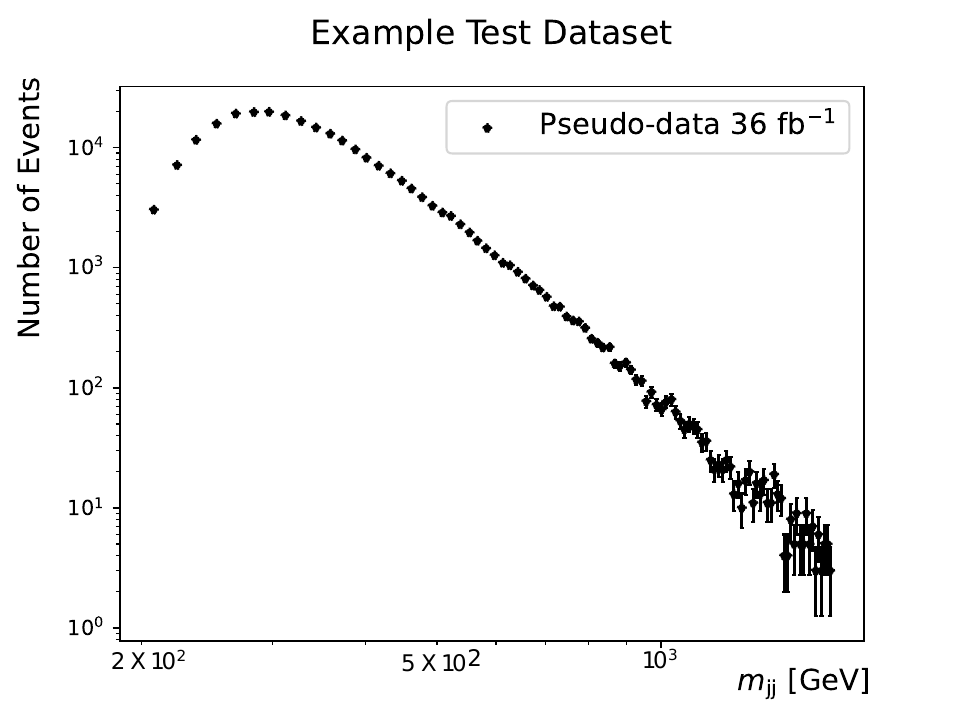}
  \caption{The smooth template used to generate the pseudo-datasets (left) and one of the example pseudo-datasets (right), corresponding to an integrated luminosity of 36 $\fb^{-1}$. \mjj refers to the invariant mass of the di-jet system.}
  \label{fig:bkg_fit_master_lhc}
  \end{center}
\end{figure}

The majority of published resonance searches consider smoothly falling backgrounds. To ensure no local features are introduced by the trigger
selection, the low-mass regions are often not used, abandoning a sizeable
amount of data. There have been attempts to include the whole dataset, such as
a search for $b$-tagged resonances conducted in CMS~\cite{CMSMultib}. In this
search, the \mjj spectrum is divided into three regions, with each region
fitted by a different function because there are no proper single functions
that can describe the entire spectrum. One clear advantage of GPR is the
ability to deal with complex spectra; hence, this work selects this analysis and
uses its published data to construct the test dataset. A large set of
pseudo-experiments is generated for the statistical tests, using the smooth
background extracted from the results presented in ref.~\cite{CMSMultib}, with
a uniform bin width of 15 GeV. Each pseudo-dataset is achieved by varying the
event count in every bin independently according to a Poisson law. The corresponding integrated luminosity is 36 $\fb^{-1}$ roughly, hereinafter referred to as the ``LHC scenario''. 
Figure~\ref{fig:bkg_fit_master_lhc} shows the seed template used to generate
the pseudo-data and one example dataset. The signal injection tests mentioned
in Section~\ref{sec:sensitivity} are generated similarly after the signal
events are added to the background. 

The binning of the dataset impacts the performance, and the choice is made
based on several confounding factors, such as the detector resolutions and
computing cost. Given the computing complexity of GPR is $\mathcal{O}(n^3)$ for
$n$ data points~\cite{GPR2006}, and the foreseeable enormous dataset expected
at the (HL-)LHC, exploring GPR using binned data is arguably more practical than considering unbinned data. A comparative study of the binning choice can offer valuable insights to the
community, although it is not studied in this work.   

\section{{\bfseries Model setup}}
\label{sec:tuning}

This work utilises the GPR module available in \SCIKITLEARN{1.5.1}, without any modifications to the core libraries. 

\subsection{Data pre-processing}
\label{sec:datapre}

\begin{figure}[ht]
    \begin{center}
    \includegraphics[width=0.4\columnwidth]{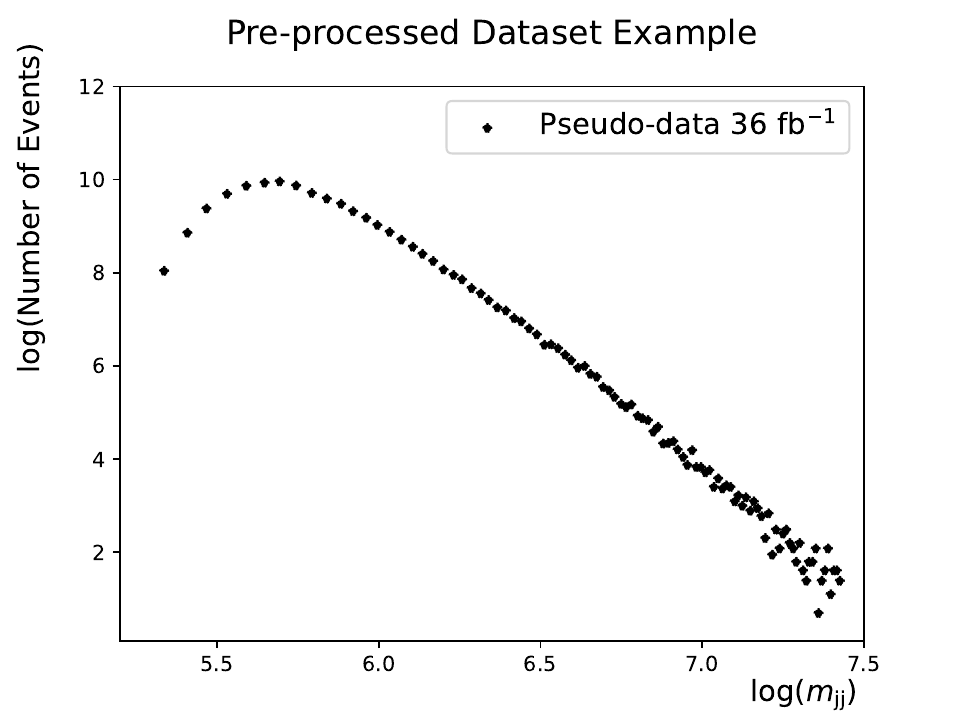}
    \includegraphics[width=0.4\columnwidth]{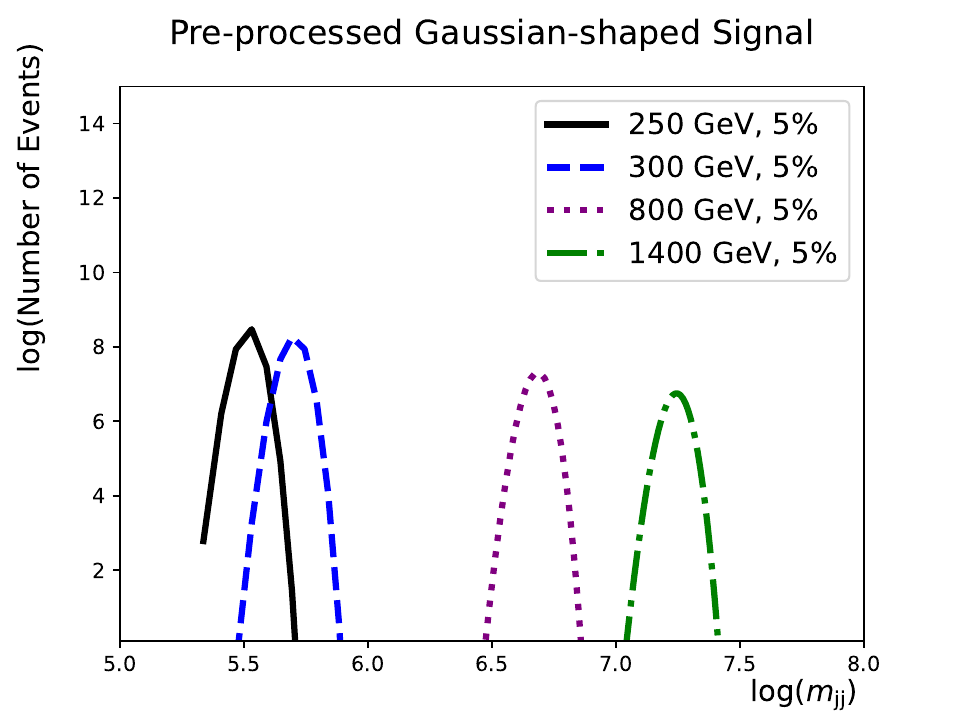}
    \caption{An example of the pre-processed background-only pseudo-datasets (left) and an illustration of the distributions for Gaussian-shaped signal at the 250, 300, 800 and 1400 GeV mass points, with a 5\% width (right). Each signal point is normalised to 10K events.}
    \label{fig:transformed_dataset}
    \end{center}
\end{figure}

The dataset used for resonance searches in the hadronic final states is often
very sizeable. In this work, we use a binned dataset, so for each bin centre
($\mjj$) there is a certain number of entries ($N_{\mathrm{events}}$),
corresponding to the input data points ($x_i$) and the ones to be predicted by
GPR ($y_i$), respectively. Its rapidly changing nature and the large mass
coverage require a dedicated pre-processing step. Applying the logarithms of
$x$ and $y$ can achieve both fast convergence and good performance. In the case
of empty bins where $\log(y_{i})$ is not defined, zero padding is adopted.
Figure~\ref{fig:transformed_dataset} shows an example pre-processed background
dataset and hypothetical Gaussian-shaped signal events. A Gaussian-shaped
signal approximates a resonance with a given mass at the mean. The width of the
resonance is parametrised by the ratio between the standard deviation and the
mean. A benchmark value of 5\% is chosen as it is a frequently considered
scenario in resonance searches. The signal is well contained within an
approximately 0.5 interval on the $x$-axis, motivating the hyperparameter
choice made in Section~\ref{sec:hyper_para}. 

\subsection{GPR model setup}
\label{sec:kernel}

The kernel choice has a significant impact on the performance. As already
mentioned in Section~\ref{sec:intro}, there is a large series of kernels to
choose from~\cite{GPR2006}. The goal of this work is not to find or construct
the most optimal kernel, but to demonstrate the generality of a GPR-based
method. Therefore, we use the multiplication of the two most common ones, an
RBF and a constant kernel: $\mathrm{C}(c)\times \mathrm{RBF}(\ell)$. Previous
works~\cite{GPR2017,GPR2022} have shown this combination is a viable choice for
HEP experiments. The only two parameters to be optimised are $c$ and $\ell$,
and the bounds for $c$ and $\ell$ are hyperparameters defined a priori. The
length scale of the RBF kernel, $\ell$, determines the smoothness of the model,
and it has to be large enough in order not to incorporate the signal events. It
is because a localised feature like a narrow signal will create bumpy features.
The prior mean of $y_i$ ($\mu_i$) has a diminishing impact on the posterior
distribution, and is set to zero. 

\subsection{Regularisation}
\label{sec:reg}

For the functional fit method, users have a set of criteria, such as what is
listed in Section~\ref{sec:validation}. When a function cannot satisfy those criteria, usually the initial reaction is to find a better-performing one. Similarly,
we can also try to craft a more suitable customised kernel in a GPR-based
strategy. However, this problem can be
addressed differently, by biasing the minimisation to meet the imposed
standards.

$L_2$ regularisation, also referred to as the Ridge regression, is a well-known
method to apply such biases~\cite{GPR2006}, which has the following general
form:

\[ \mathbf{K} \rightarrow \mathbf{K} + \lambda \mathbf{I} \]

where $\mathbf{K}$ is the covariance matrix, $\lambda$ is a scalar and
$\mathbf{I}$ is a diagonal matrix. The biasing term $\lambda \mathbf{I}$ can be regarded
as the level of noise expected in the training data. It should not be
homoscedastic as the relative uncertainties change across the spectrum.
Therefore, a diagonal matrix with varying diagonal elements is used instead of
$\lambda \mathbf{I}$. The implementation is realised in \textsc{Scikit-learn}
via an array of those diagonal elements \footnote{The $\alpha$ parameter in the
GaussianProcessRegressor module.}, effectively a set of hyperparameters to
tune. The nominal value is set to $\alpha_i = \sqrt{y_i}/(y_i\log{y_i})$, which
is the relative uncertainty of $\log{y_i}$ propagated from $y_i$. It gives much
better performance and faster convergence than using the absolute uncertainty
of $\log{y_i}$. For the zero-padded $\log{y_i}$, $\alpha_i$ is set to unity. As
discussed later, in the low mass region where the turning point of the spectrum
lies, the corresponding $\alpha_i$ can be decreased to improve the performance. 

\subsection{Hyperparameters}
\label{sec:hyper_para}    
 
Table~\ref{tab:hyperparameters} summarises the hyperparameters and their
nominal values that are fixed during the optimisation of $c$ and $\ell$. The lower bound on $\ell$ is set to 0.5, motivated by Figure~\ref{fig:transformed_dataset}. The upper bound is set to 20, which is
more than three times larger than the width of the spectrum. The bounds on the
constant kernel ($c$) are set to an arbitrarily large or small number, which makes it effectively unconstrained. The tuning of the $L_2$ regularisation terms is achieved by
a set of multiplication factors ($f_i$) applied to $\alpha_i$ to satisfy the performance criteria, as discussed in Section~\ref{sec:validation}.    

\begin{table}[htbp]
  \begin{center}
    \caption{The list of hyperparameters of the GPR model.}
    \makebox[0pt]{
\renewcommand{\arraystretch}{1.2}
\begin{tabular}{|c|c|c|}
\hline
 Name & Explanation & Nominal Values \\
\hline
$\mathbf{\ell}_{0}$ & RBF kernel length scale lower bound & 0.5 \\ 
$\mathbf{\ell}_{1}$ & RBF kernel length scale higher bound & 20 \\ 
$\mathbf{c}_{0}$ & constant kernel lower bound & $10^{-5}$ \\ 
$\mathbf{c}_{1}$ & constant kernel higher bound & $10^{18}$ \\ 
$\mathbf{\alpha}_{i}$ & diagonal elements added to the kernel matrix & $\sqrt{y_i}/(y_i\log{y_i})$  \\ 
$\mathbf{f}_{i}$ & multiplication factors applied to $\mathbf{\alpha}_{i}$ & 1  \\ 
\hline
\end{tabular}
}

    \label{tab:hyperparameters}
  \end{center}
\end{table} 

\section{{\bfseries Background estimate validation}}
\label{sec:validation}

A background modelling method should be validated with background-only samples.
Well-known goodness-of-fit tests, such as the reduced $\chi^2$
test~\cite{chi2test}, can quantify the modelling performance, where a
$\chi^2/\mathrm{nDoF}$ close to unity indicates a well-behaved background
estimate. In addition, the Kolmogorov--Smirnov (KS) test evaluates how
compatible the significance is with a normal distribution~\cite{kstest}, as it is
expected when the background estimate is unbiased. A KS $p$-value smaller than
the threshold rejects the hypothesis that the significance is a normal distribution.
An ensemble test using pseudo-experiments tells us the fraction of trials that
gives an acceptable reduced $\chi^2$ or KS test result. Though these two classic
tests reveal the quality of the overall performance, they cannot detect local
biases efficiently. As the search strategy is designed to find narrow peaks,
such mis-modelling is likely to induce false-positive errors. 

The false-positive rate depends on the specific statistical test
adopted by the analysis. In this study, we consider a model-agnostic method
named ``Bump Hunter'' (BH), which calculates the probability for the largest
deviation between data and background estimate to originate from statistical
fluctuations. The corresponding mass interval is also identified. A BH
$p$-value smaller than the threshold rejects the hypothesis that the largest
deviation is due to background fluctuations~\cite{BH}, which is a false-positive if a background-only spectrum is analysed. Similarly, pseudo-experiments are performed to assess the false-positive rate. The BH
$p$-values reported from pseudo-experiments considering background-only test
datasets are expected to be flat if no obvious local biases are introduced in the background modelling. A \textsc{python} implementation of this algorithm,
\textsc{pyBumpHunter}~\cite{pyBumpHunter}, is used.

\begin{figure}[ht]
  \begin{center}
   \includegraphics[width=0.4\columnwidth]{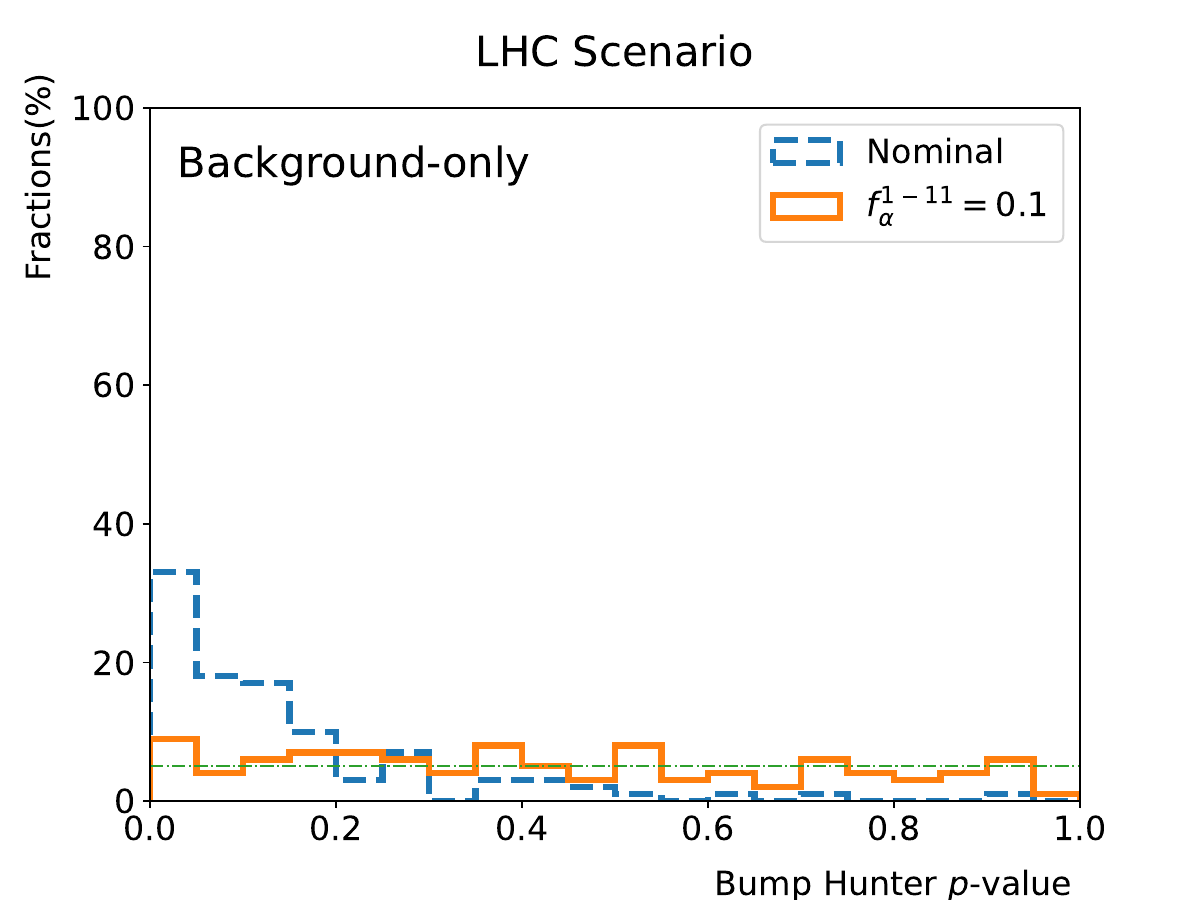}
  \caption{BH $p$-value distributions of 100 background-only pseudo-experiments in the LHC scenario, for the nominal setup (dashed line) and $f_{1-11} = 0.1$ (solid line). The green dashed line indicates the ideal distribution expected without injected signal events.}
  \label{fig:bh_bkg_only}
  \end{center}
\end{figure}

\begin{figure}[ht]
  \begin{center}
   \includegraphics[width=0.4\columnwidth]{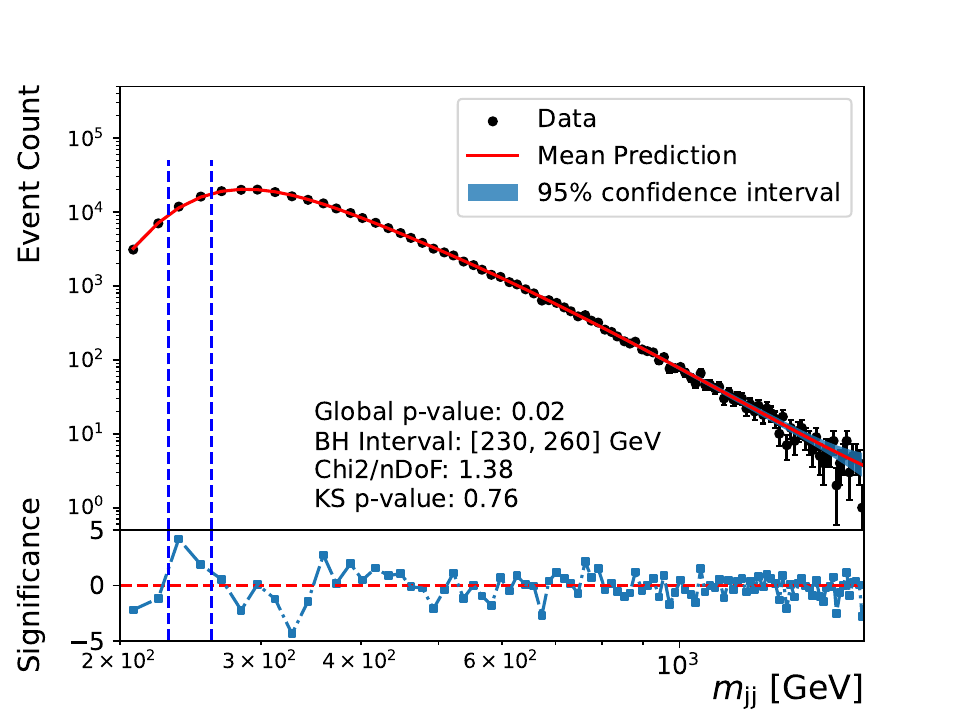}
   \includegraphics[width=0.4\columnwidth]{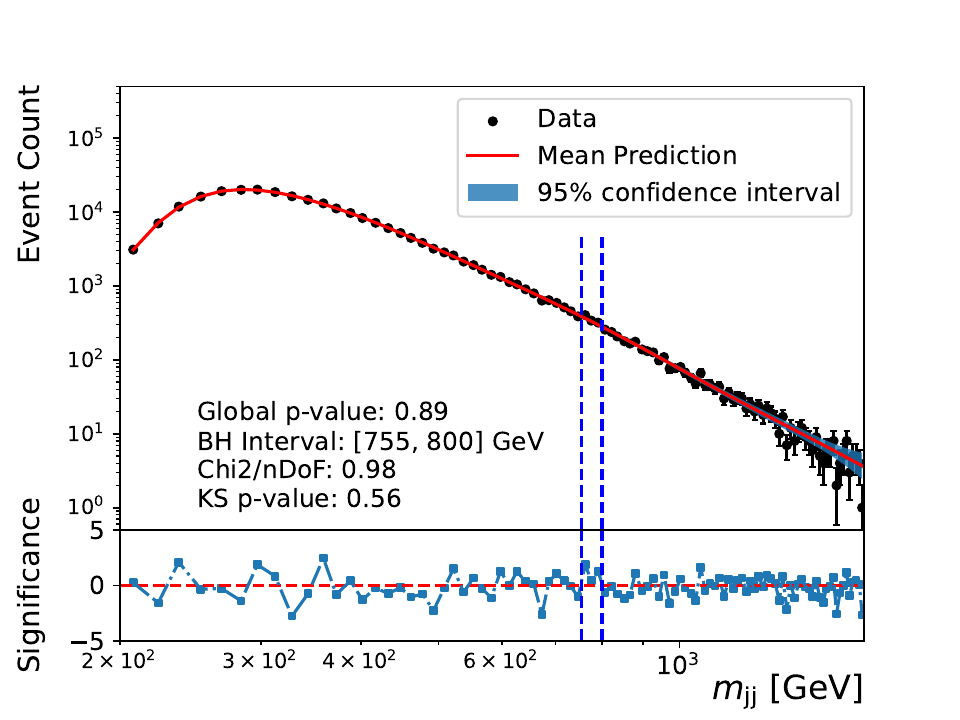}
  \caption{Comparison between an example background-only pseudo-dataset (solid point) and the background estimate from GPR (solid line), for the nominal setup (left) and $f_{1-11} = 0.1$ (right), in the LHC scenario. The vertical dashed lines indicate the boundaries of the most significant deviation reported by BH. The lower panel shows the significance calculated for each mass bin.}
  \label{fig:bkg_only_fit_example}
  \end{center}
\end{figure}

It is found that shrinking $f_i$ associated with the mass bins before the
smoothly falling part improves the precision in the corresponding \mjj
region. A coarse scan using 5 random pseudo-experiments indicates that once
$f_{i}$ below 320 GeV is reduced to 0.1 from unity, the performance becomes
stable. Table~\ref{tab:fit_validation} summarises the test results from the
nominal setup and the one where the first eleven $f_i$ parameters are set to 0.1
($f_{1-11} = 0.1$), covering the mass range up to 365 GeV. It is a
configuration, randomly selected among the tested ones, where up to the first
fifteen $f_{i}$ parameters (\mjj up to 425 GeV) are scaled. The latter case achieves much
better performance in all three tests. Furthermore, as seen in
Figure~\ref{fig:bh_bkg_only}, the BH $p$-values are homogeneously distributed
as expected. Figure~\ref{fig:bkg_only_fit_example} illustrates the performance
difference in one of the pseudo-experiments. 

\begin{table}[htbp]
    \begin{center}
    \caption{Summary of the fit performance for the nominal $f_i$ and setting $f_{1-11}$ to 0.1. The fractions are calculated based on 100 pseudo-experiments, so the uncertainties vary from 3\% to 5\%, assuming binomial errors. Pseudo-experiments satisfying the criteria are considered successful.}
    \makebox[0pt]{
\renewcommand{\arraystretch}{1.2}
\begin{tabular}{|c|c|c|}
\hline
Criteria  & Nominal & $f_{\alpha}^{1-11} = 0.1$ \\
\hline
KS p-value > 0.05 & 86\% & 88\%\\ 
$\chi^2$/nDoF < 1.5 & 65\% & 85\%\\ 
BH p-value > 0.1 & 51\% & 87\%\\ 
\hline
\end{tabular}
}

    \label{tab:fit_validation}
   \end{center}
\end{table}

\begin{figure}[ht]
  \begin{center}
   \includegraphics[width=0.4\columnwidth]{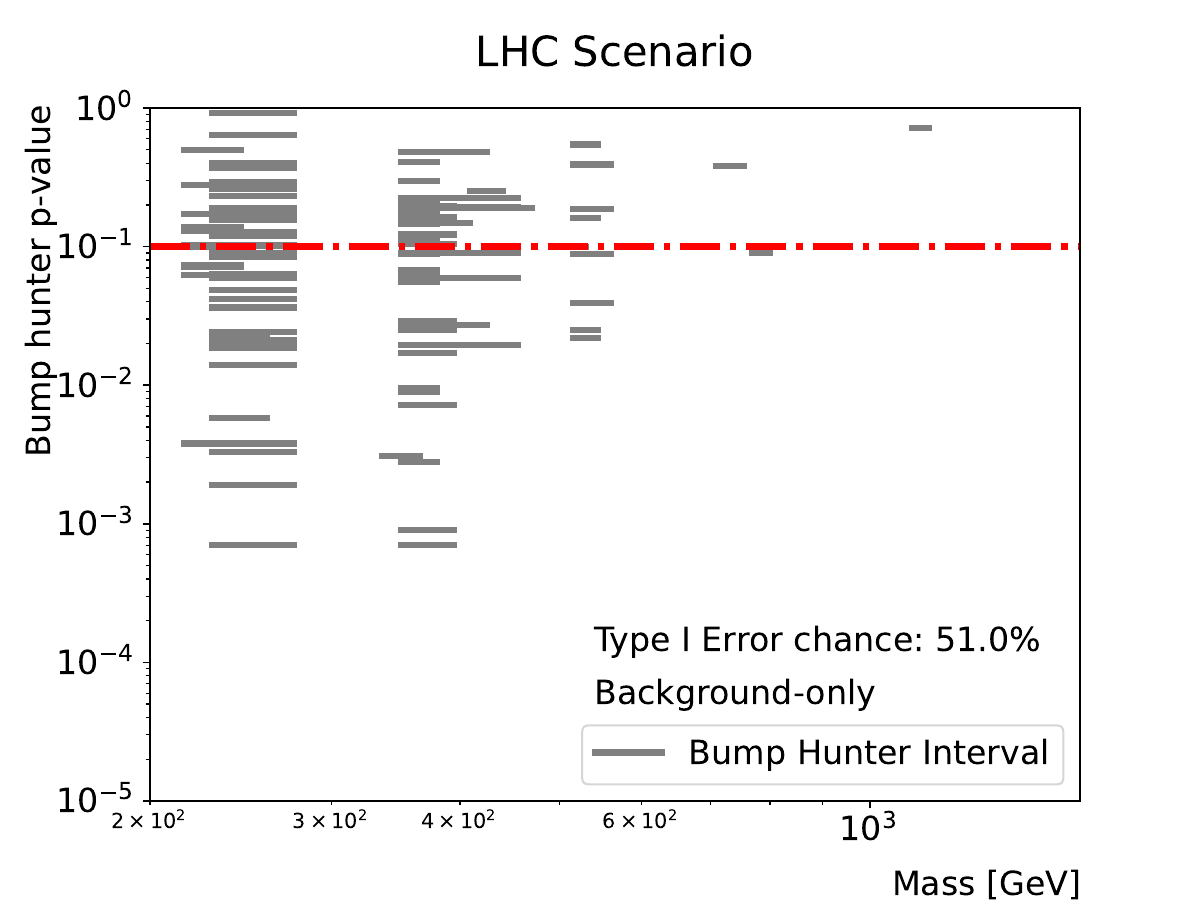}
   \includegraphics[width=0.4\columnwidth]{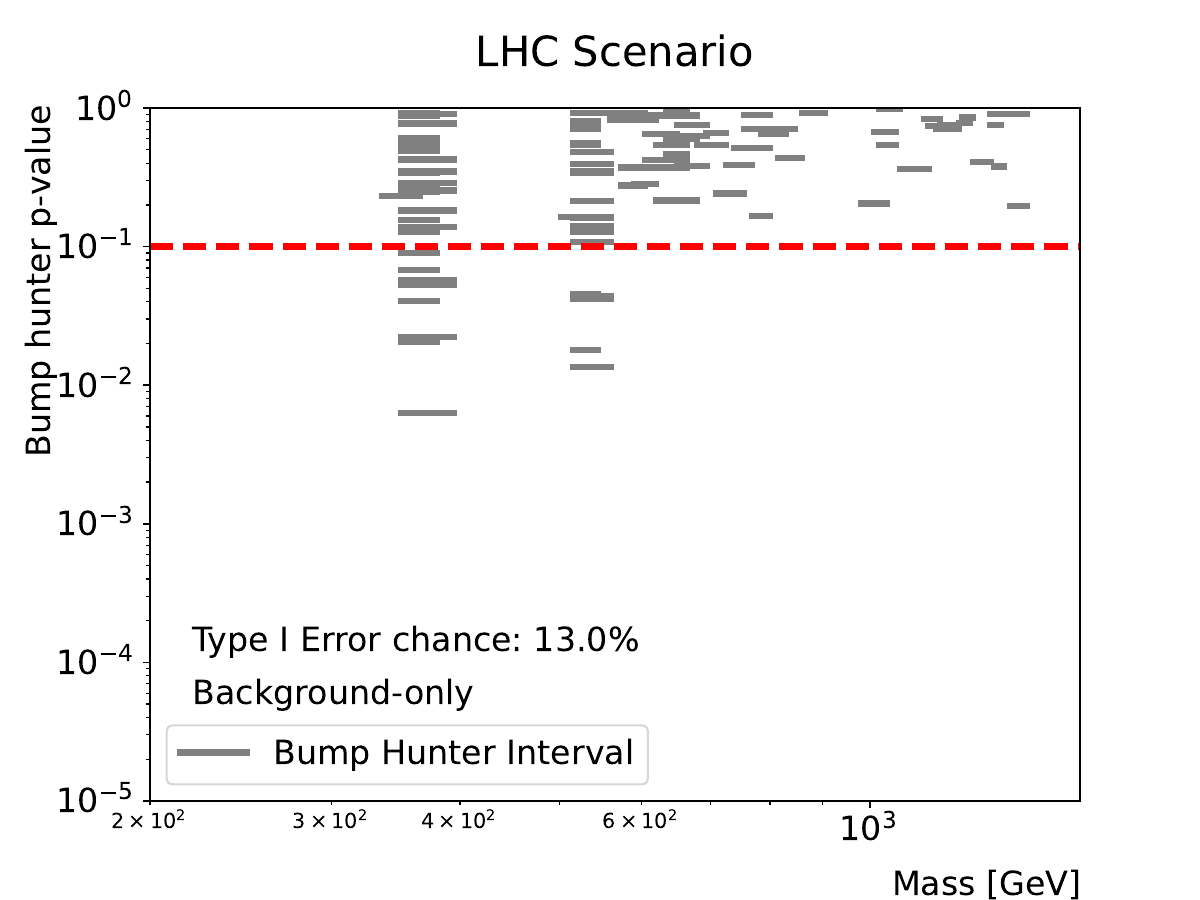}
  \caption{Summary of the BH $p$-values and the corresponding mass intervals (solid horizontal segments), for the nominal setup (left) and $f_{1-11} = 0.1$ (right), in the LHC scenario. Each solid horizontal segment comes from one pseudo-experiment. The horizontal dashed line corresponds to a critical value of 0.1. Flagged intervals with BH $p$-values above this threshold are considered not significant.}
  \label{fig:bh_edges_bkg_only}
  \end{center}
\end{figure}

Though the background modelling has passed the above tests, there can still be
remaining biases that need to be considered as the uncertainties. Such
residual biases are observed in Figure~\ref{fig:bh_edges_bkg_only}, summarising
the most significant BH intervals. The predominant cluster near 250 GeV in the
nominal case is mitigated largely by changing $f_{1-11}$ to 0.1, but those two
clusters localised near 400 and 500 GeV are still visible~\footnote{These systematic effects are in part due to the manual extraction of background from ref~\cite{CMSMultib}}. There is plenty of
freedom to tune $f_{i}$ further, but the rest of the work uses this benchmark
setup.

\section{{\bfseries Sensitivity study}}
\label{sec:sensitivity}

Previously, we have identified a configuration that gives us acceptable
background modelling precision in the background-only case. This section will
demonstrate that this configuration retains good sensitivity to signals across
the entire spectrum. When it comes to analyses using data-driven background
estimate methods, we have to check carefully how sensitive the method is to the
possible signal events. It is evaluated by a series of signal injection tests
with Gaussian-shaped signal events. Four mass points are included to cover
various locations in the spectrum. The 250 \GeV point is below the trigger
threshold, and the 300 \GeV point is right at the plateau. These regions are
often discarded in physics analyses. The 1400 \GeV point is close to the end of
the \mjj distribution, while the 800 \GeV point is in the middle of the
smoothly falling region where optimal sensitivity is expected when a
traditional functional fit strategy is applied. Appendix~\ref{app:sig_inj}
discusses those various injection cases in further detail. 

The amount of signal events injected is quantified by $s/\sqrt{b}$ in a given
\mjj window centred at the signal mass, covering 68.3\% of the signal events.
Table~\ref{tab:injection} lists the injection tests conducted. One hundred test
datasets are prepared for each mass point and then fed into the sensitivity
evaluation procedure as detailed in the next section.   

\begin{table}[htbp]
   \begin{center}
   \caption{Summary of the signal injection tests. The signal strength is defined using the unit of $s/\sqrt{b}$.}
   \makebox[0pt]{
\renewcommand{\arraystretch}{1.2}
\begin{tabular}{|c|c|c|}
\hline
Mass [GeV] & Width & Strength\\
\hline
250 & 5\% & 10 and 15\\ 
300 & 5\% & 7 and 10\\
800 & 5\% & 5 and 7 \\
1400 & 5\% & 5 and 7 \\
\hline
\end{tabular}
}

   \label{tab:injection}
  \end{center}
\end{table}     

\subsection {Sensitivity evaluation}
\label{sec:bh_results}

\begin{figure}[ht]
  \begin{center}
   \includegraphics[width=0.35\columnwidth]{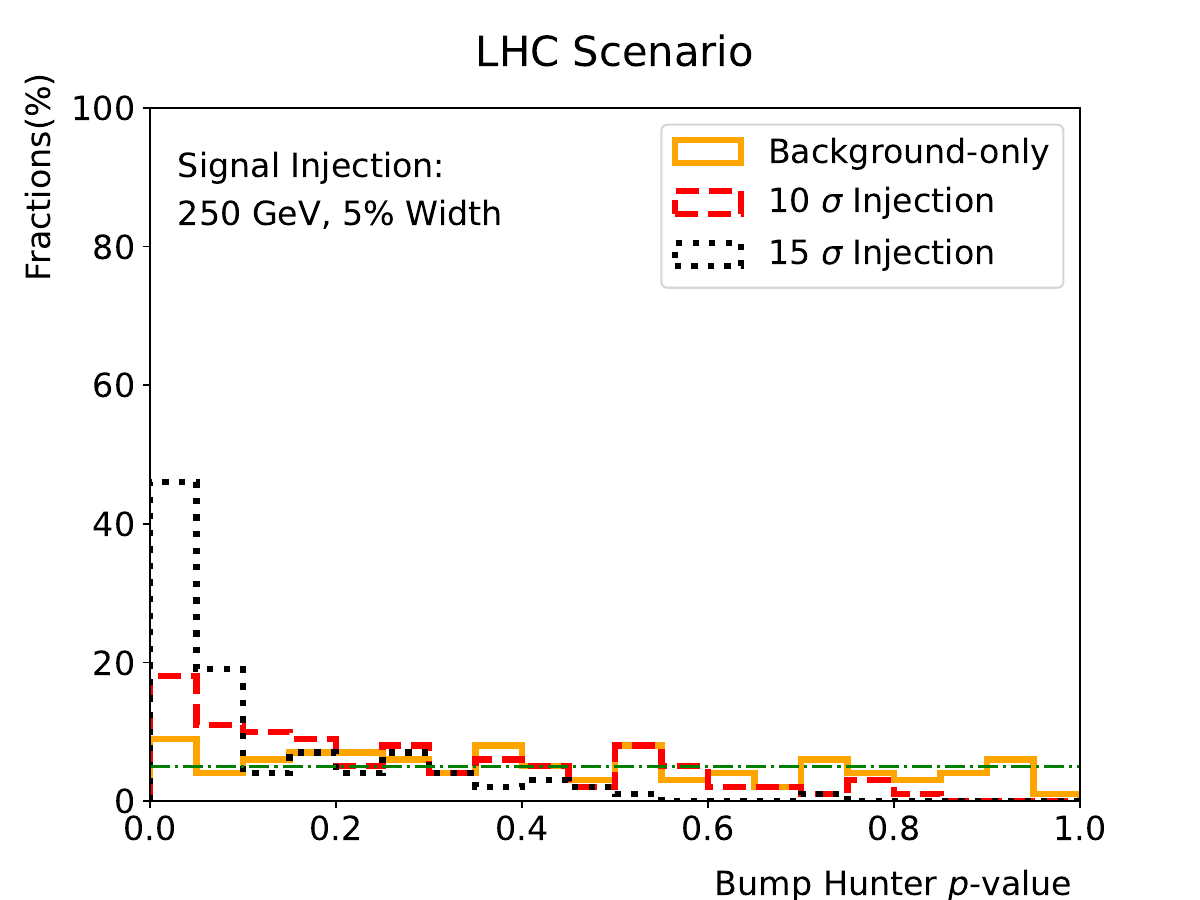}
   \includegraphics[width=0.35\columnwidth]{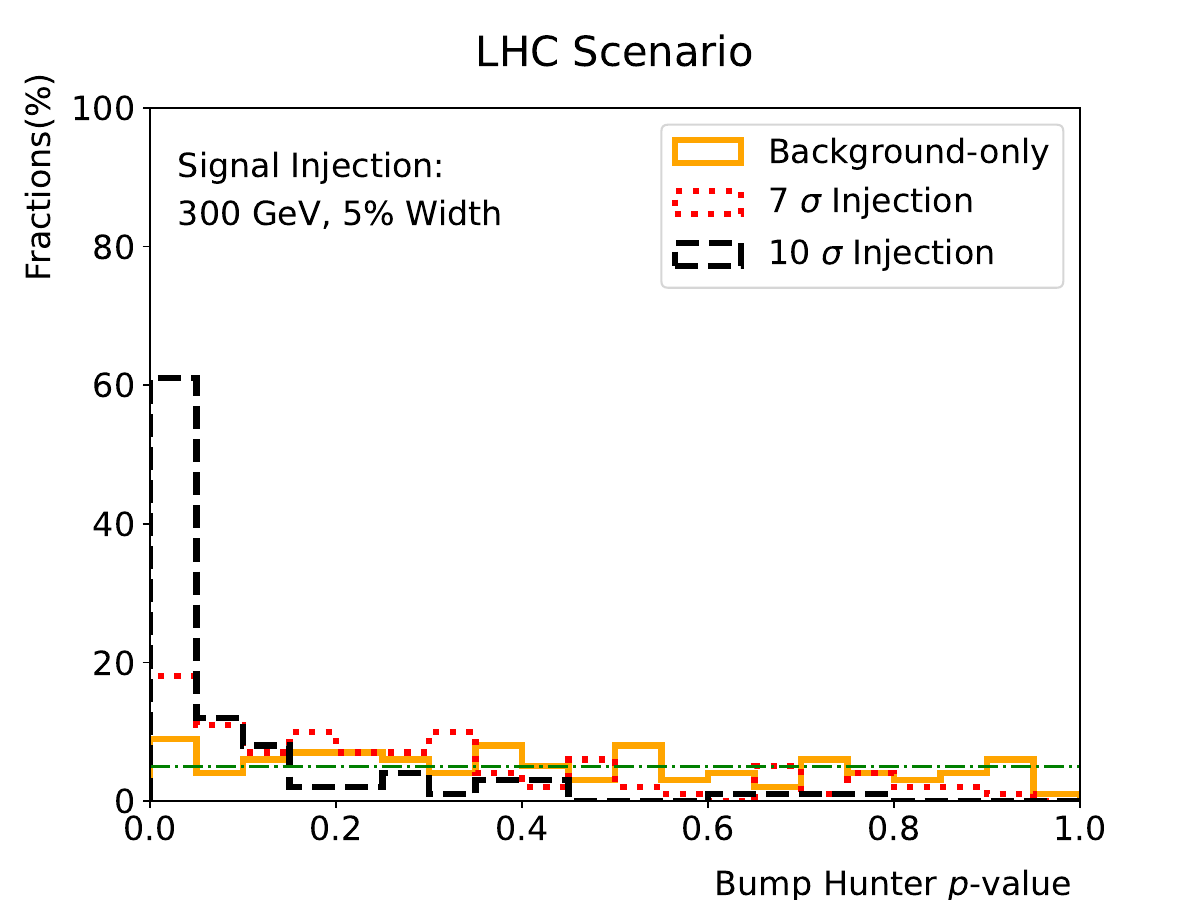}
   \includegraphics[width=0.35\columnwidth]{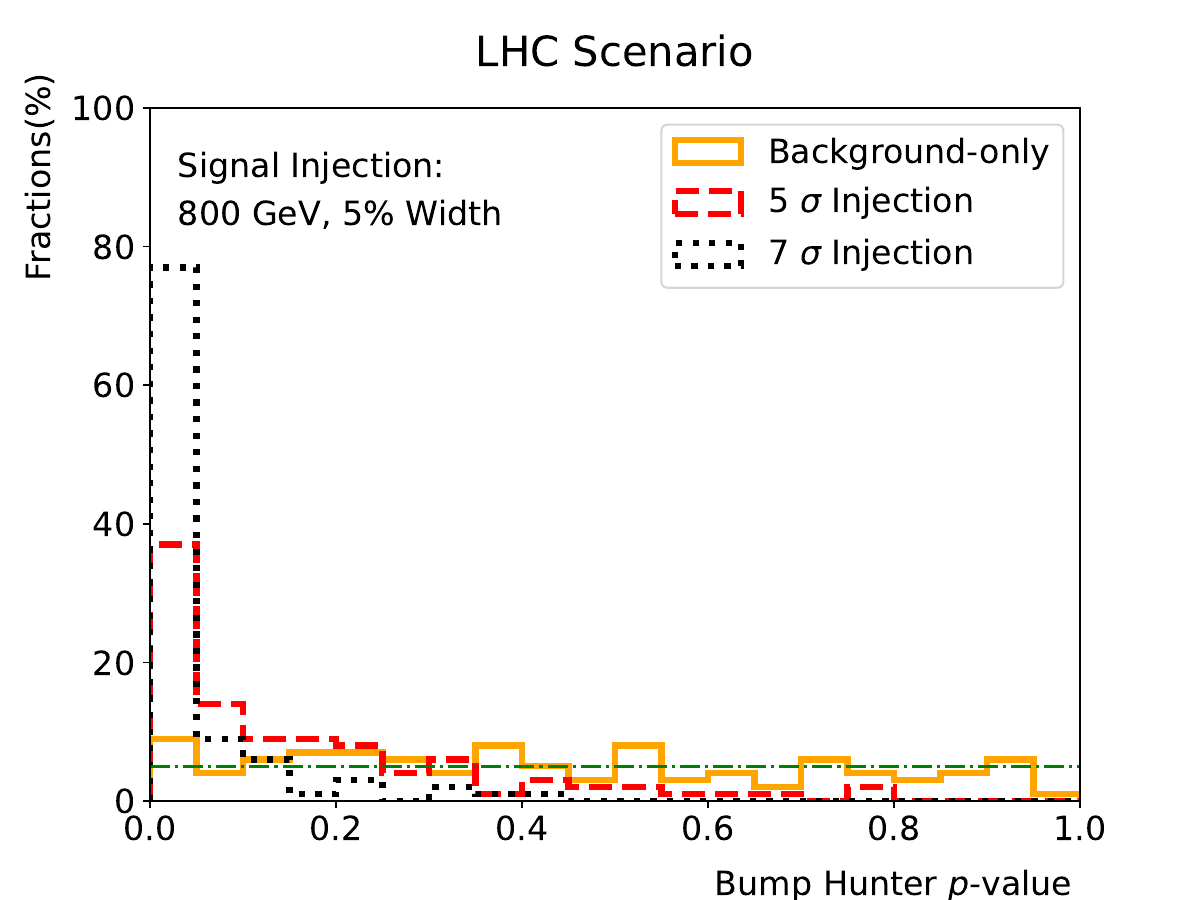}
   \includegraphics[width=0.35\columnwidth]{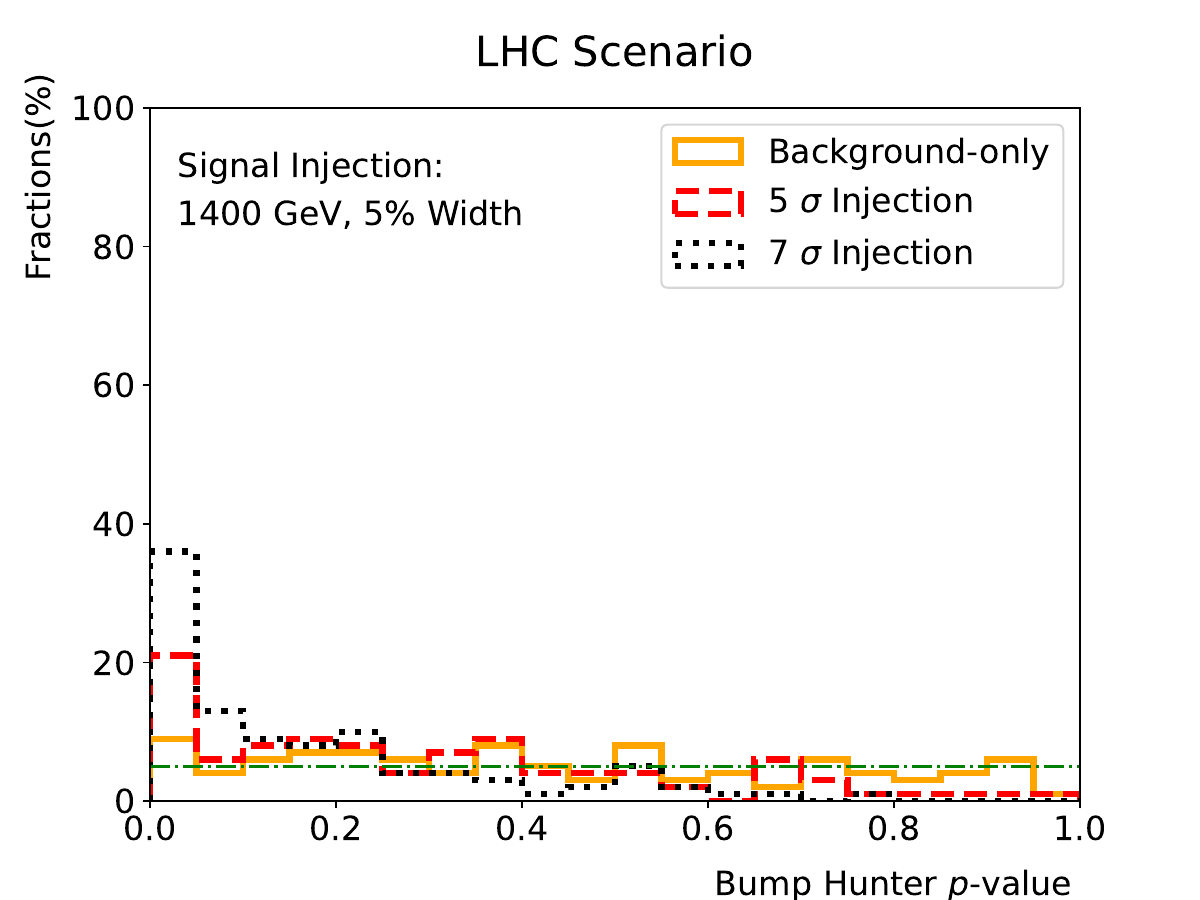}
  \caption{Summary of BH $p$-values obtained in the signal injection tests, for the 250 GeV (upper-left), 300 GeV (upper-right), 800 GeV (lower-left) and 1400 GeV (lower-right) signal mass points, in the LHC scenario. Each test consists of 100 signal-injected pseudo-experiments. The green dashed line indicates the expected distribution when no signal events are injected.}
  \label{fig:injection_p_values}
  \end{center}
\end{figure}

The sensitivity of a search should be evaluated with the full analysis chain
executed. There exist several metrics to quantify the sensitivity, such as
the expected exclusion limits for given BSM hypotheses. Both the CMS and ATLAS
analyses often offer model-independent results using the BH algorithm, which
reports a global $p$-value indicating how likely the most significant deviation
observed comes from background fluctuations~\cite{BH}. The signal injection
test can be done using pseudo-experiments where the probability of reporting a
$p$-value below the threshold is examined for a given amount of signal events
injected. The $p$-value threshold is chosen to be 0.1 in the signal injection
test as well. As shown in Figure~\ref{fig:injection_p_values}, the sensitivity
is optimal for the region in the middle with sizeable sidebands on both sides to
constrain the background estimate, and it drops when moving towards the end of
the spectrum. The performance is reduced even more in the low mass region below
(on) the plateau.  
 
\begin{figure}[ht]
  \begin{center}
   \includegraphics[width=0.35\columnwidth]{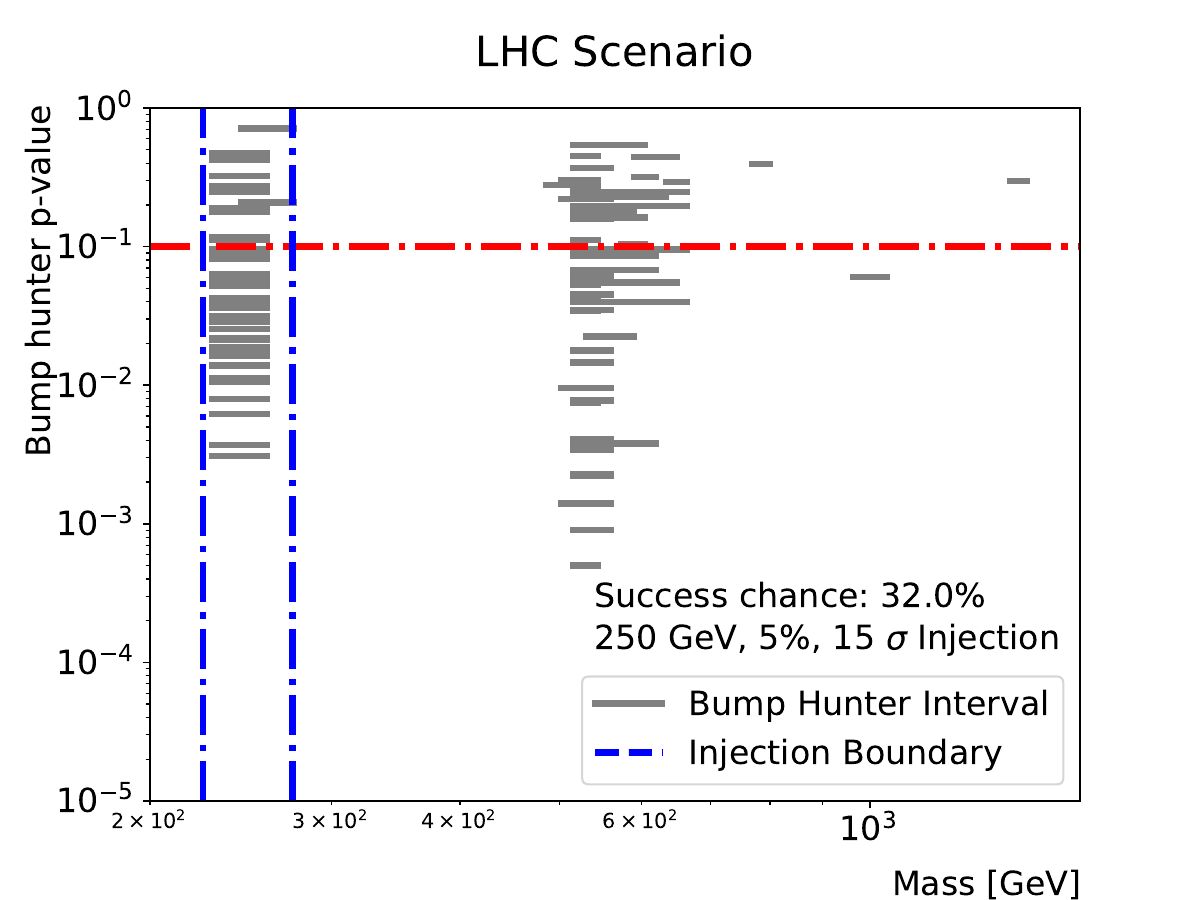}
   \includegraphics[width=0.35\columnwidth]{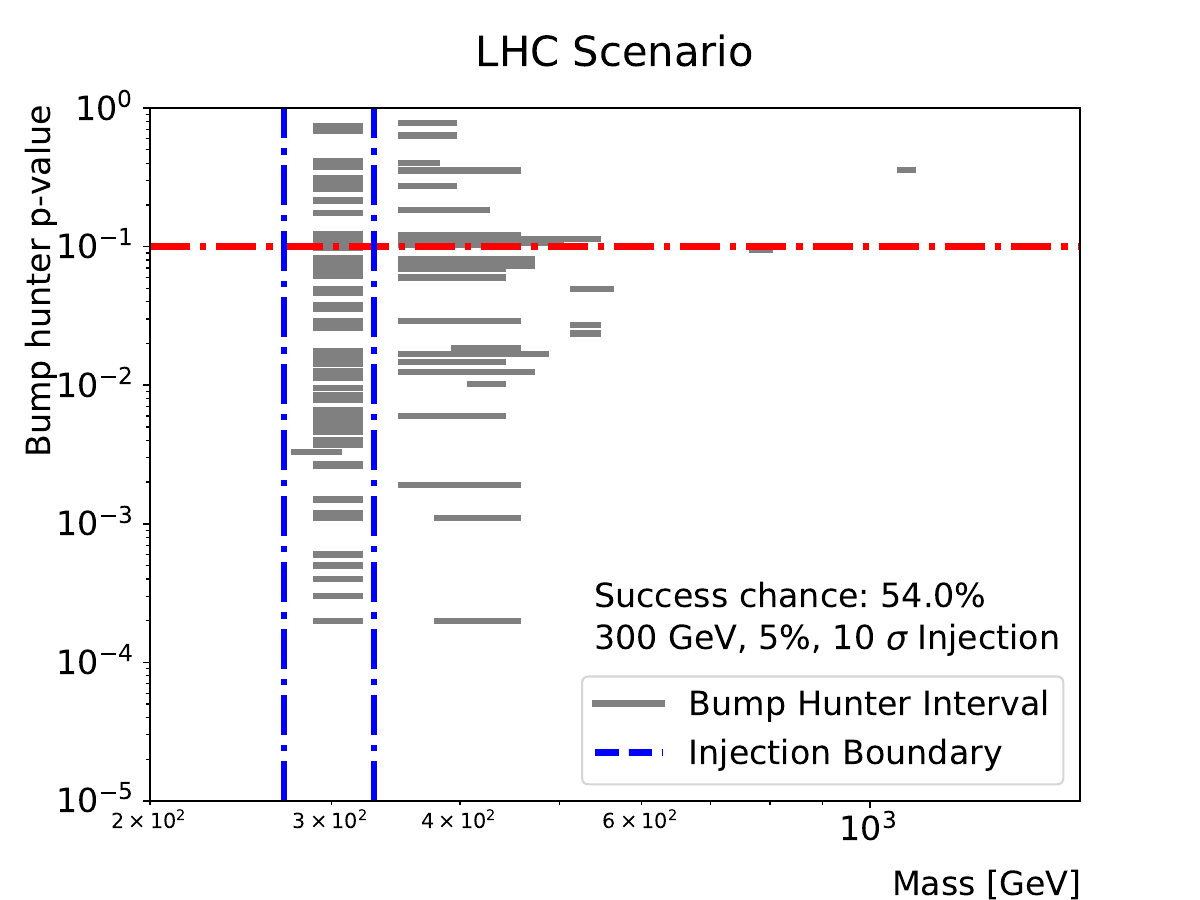}
   \includegraphics[width=0.35\columnwidth]{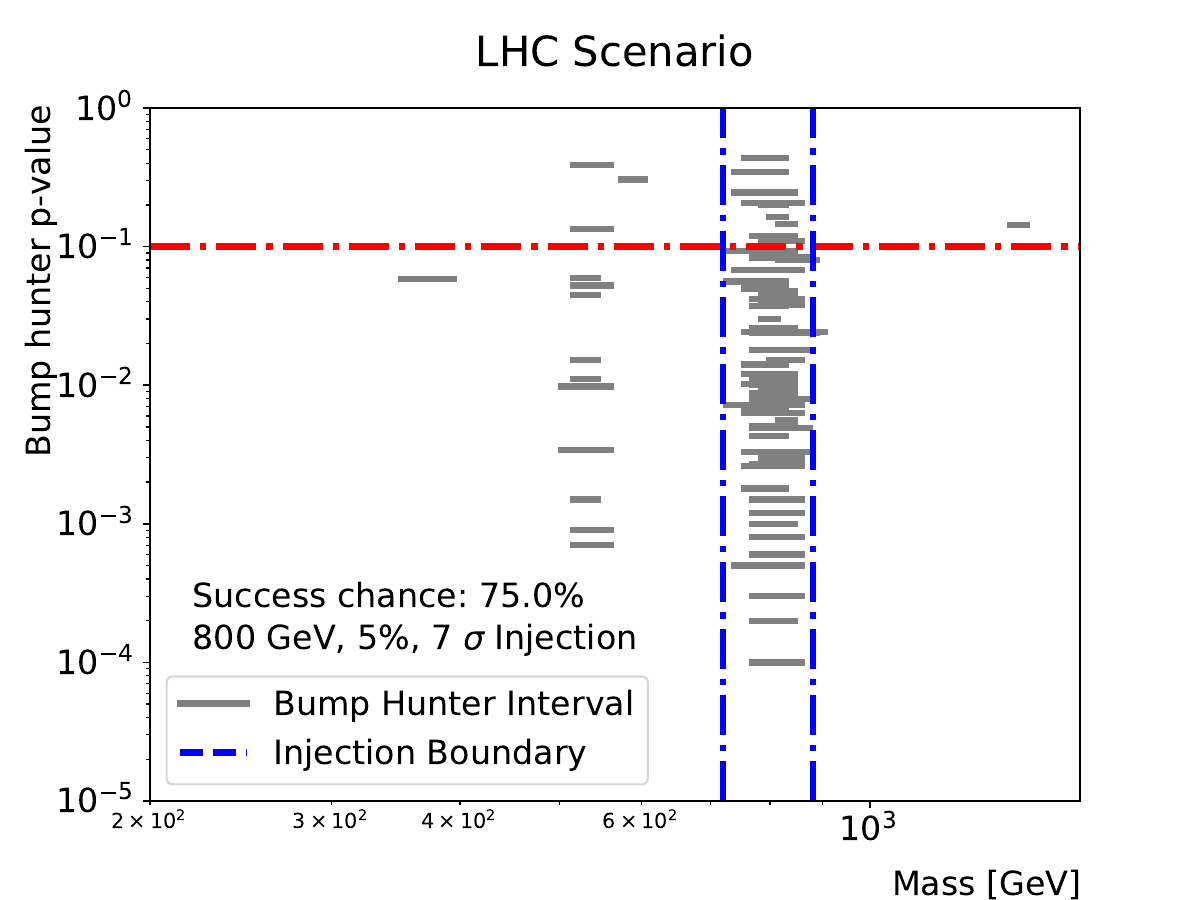}
   \includegraphics[width=0.35\columnwidth]{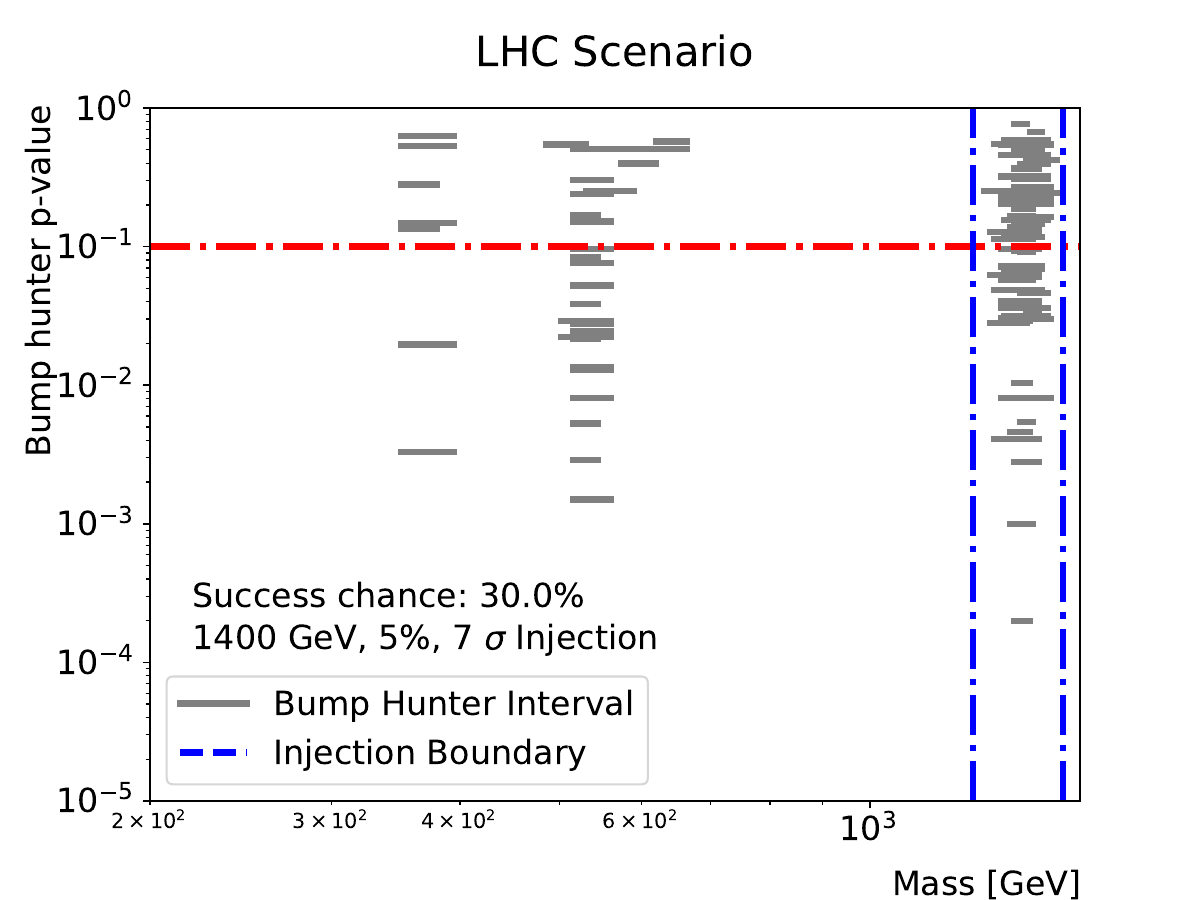}
  \caption{Summary of the BH $p$-values and the corresponding mass intervals (solid horizontal segment), for the 250 GeV (upper-left), 300 GeV (upper-right), 800 GeV (lower-left) and 1400 GeV (lower-right) signal mass points, in the LHC scenario. Each solid horizontal segment comes from one pseudo-experiment. The horizontal dashed line corresponds to a critical value of 0.1. Flagged intervals with BH $p$-values above this threshold are considered not significant. The vertical dashed-dotted lines represent the \mjj region where signal events are injected.}
  \label{fig:injection_p_values_edges}
  \end{center}
\end{figure}

Figure~\ref{fig:injection_p_values_edges} presents the chance of successfully
reporting a $p$-value less than 0.1 in the correct mass region where the signal
events are injected. It is noticed that when injecting a 300 GeV signal, there
is a high chance of reporting a significant deviation at the wrong location. It
is related to the residual bias near 400 GeV seen in
Figure~\ref{fig:bh_edges_bkg_only}. Such mis-modelling effects should be taken
into account as systematic uncertainties, or resolved via further tuning of the
model parameters. Figure~\ref{fig:injection_p_values_hl_lhc} in
Section~\ref{sec:hllhc} observes a similar but predominating effect, which is
mitigated by updated multiplication factors ($f_i$) as seen in
Appendix~\ref{app:hllhc_tuning}. Figure~\ref{fig:injection_examples} shows
random examples of the pseudo-experiments that have successfully reported a
$p$-value below 0.1, with the BH intervals marked as well. 

\begin{figure}[ht]
  \begin{center}
   \includegraphics[width=0.35\columnwidth]{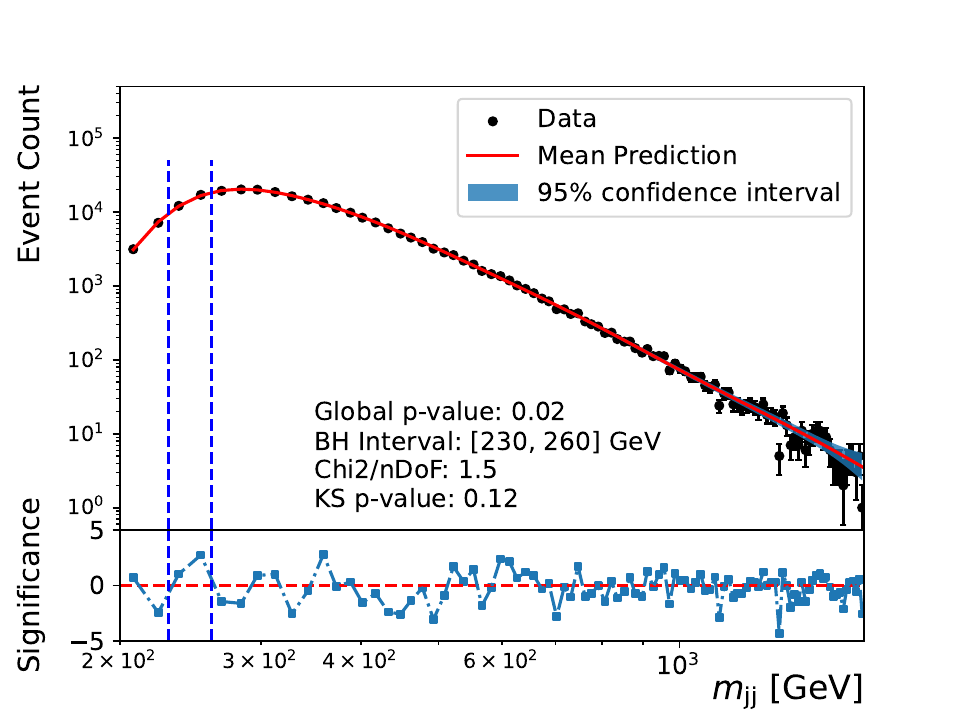}
   \includegraphics[width=0.35\columnwidth]{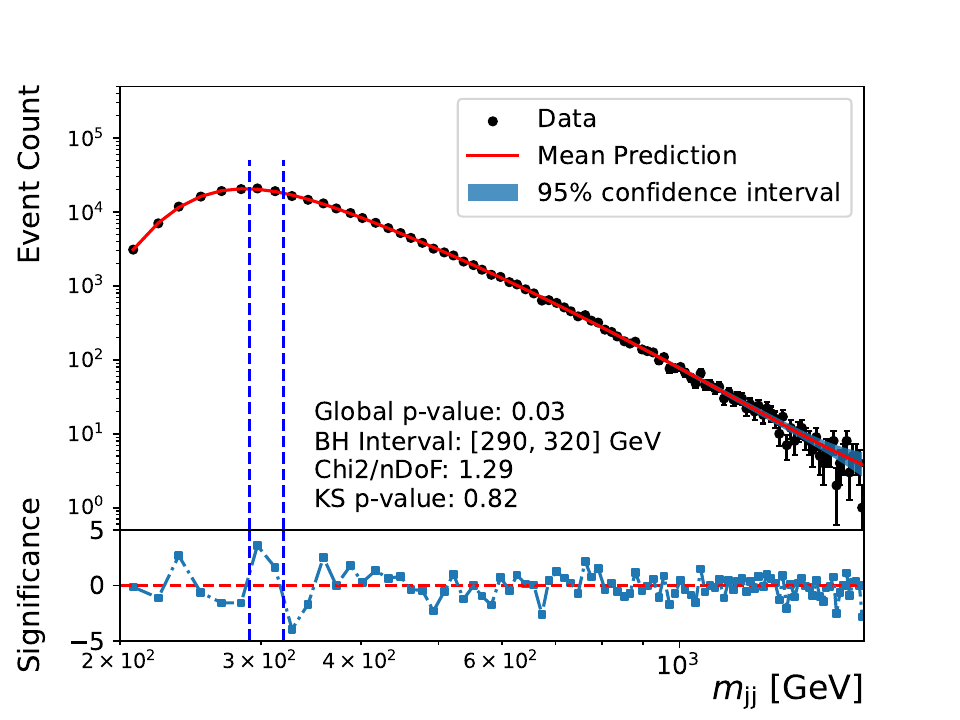}
   \includegraphics[width=0.35\columnwidth]{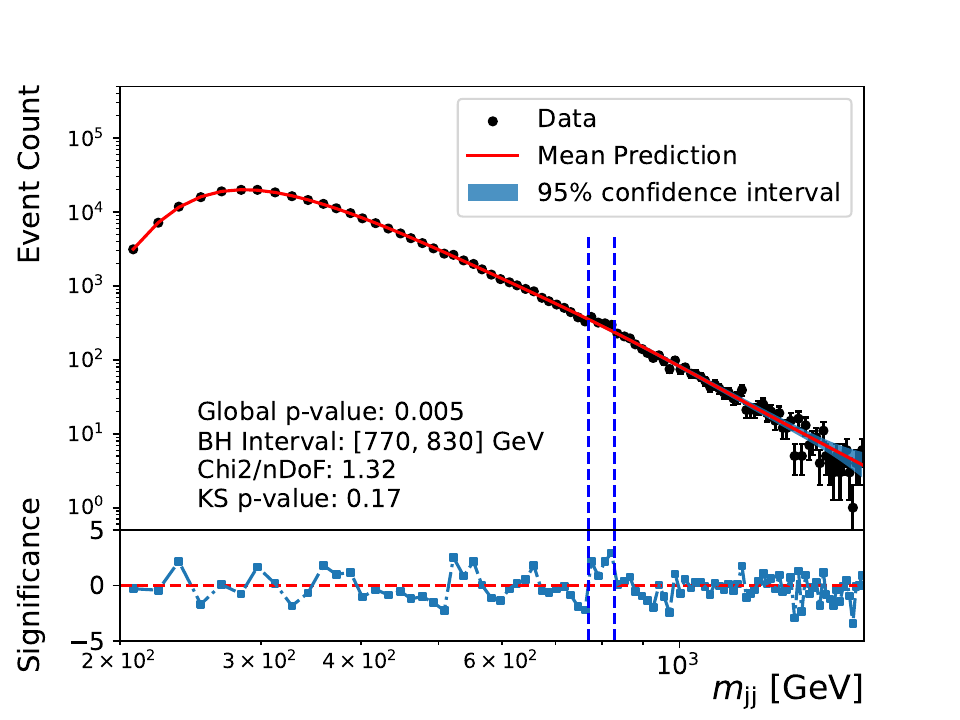}
   \includegraphics[width=0.35\columnwidth]{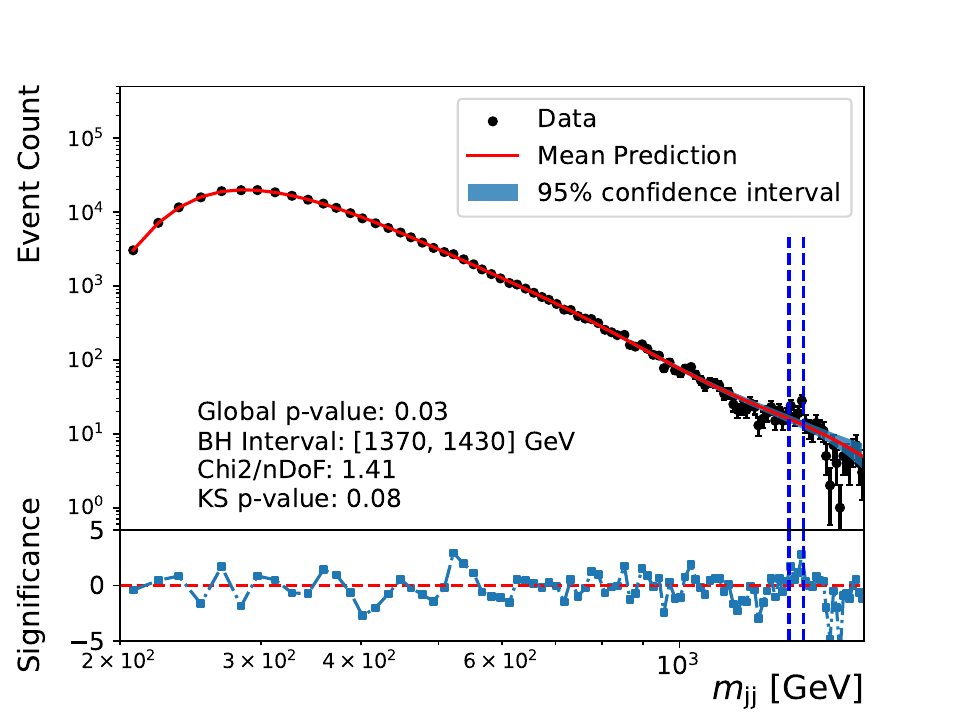}
  \caption{Comparison between the signal-injected pseudo-dataset (solid point) and the background estimate from GPR (solid line), for the 250 GeV (upper-left), 300 GeV (upper-right), 800 GeV (lower-left) and 1400 GeV (lower-right) signal mass points, in the LHC scenario. The vertical dashed lines indicate the boundaries of the most significant deviation reported by BH. The bottom panels present the significance calculated for each mass bin.}
  \label{fig:injection_examples}
  \end{center}
\end{figure}    

The 250 GeV signal point is very close to the starting of the mass spectrum,
with no constraints from the sideband on the low mass side, so the
sensitivity is naturally degraded. Figure~\ref{fig:injection_p_values} reveals
that the probability of reporting $p$-values below 0.1 only starts to increase
visibly when more than ten times of $s/\sqrt{b}$ signal events are injected.
Though this behaviour appears to be not ideal, it enables analysing a region often
discarded in physics analyses. As demonstrated in this work, it is possible to
probe the entire mass region via a unified approach. 

\begin{figure}[ht]
  \begin{center}
   \includegraphics[width=0.45\columnwidth]{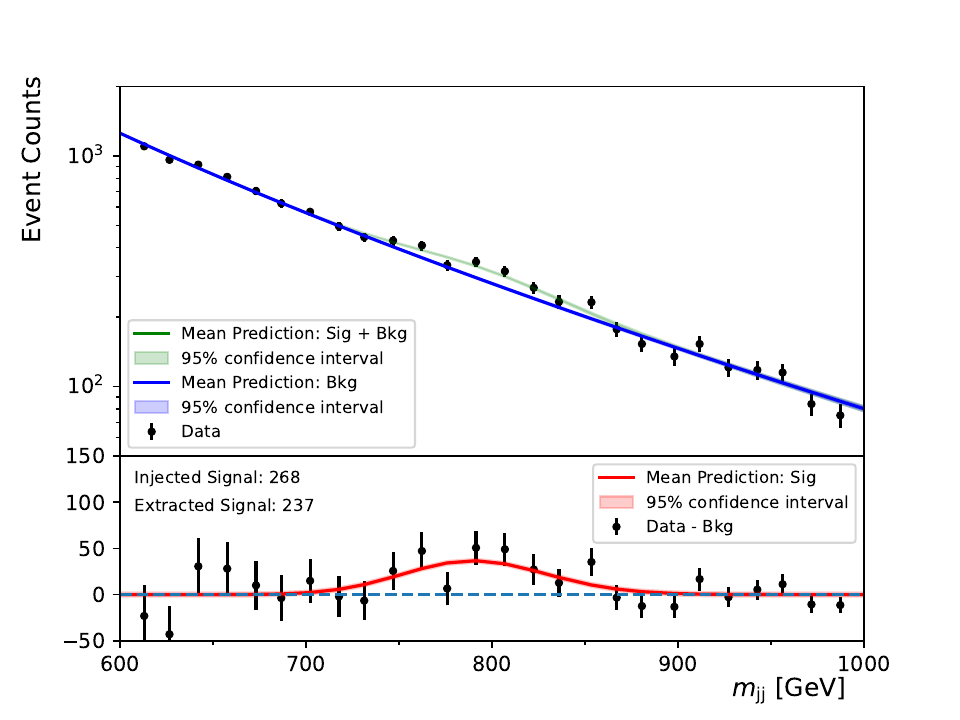}
   \includegraphics[width=0.45\columnwidth]{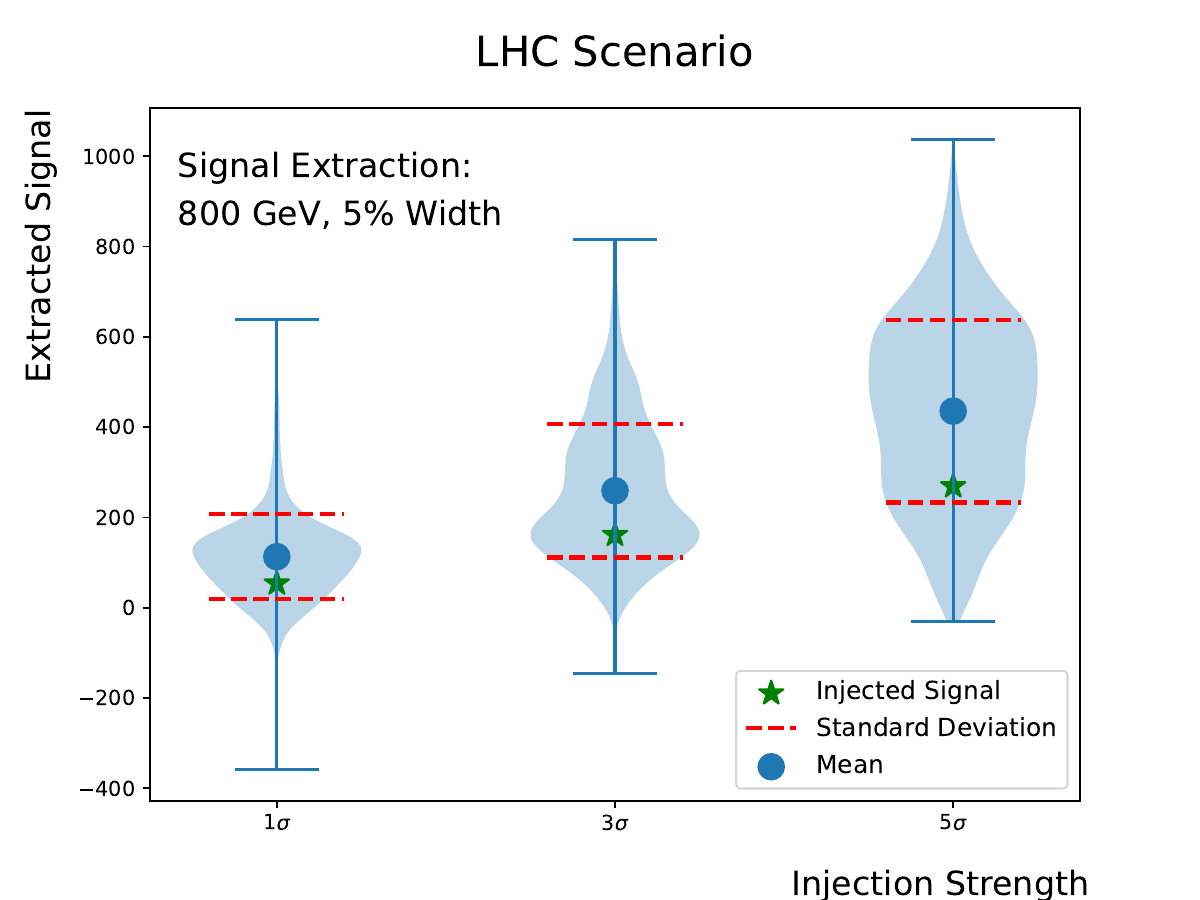}
  \caption{An example of signal extraction test done for the 800 GeV Gaussian-shaped signal with a 5\% width (left), and the summary of test results with different injected signal strengths (right), in the LHC scenario. The numbers of injected signal events are indicated by the green stars, while the blue dots represent the means of extracted signal. The width of the shaded bands corresponds to the density of a given number of extracted signal events. The red dotted lines are the one standard deviation boundaries.}
  \label{fig:extraction_examples}
  \end{center}
\end{figure}

The above tests do not rely on any signal hypotheses, as the statistical
analysis aims at identifying significant deviations in data without analysing
the nature of the deviations or quantifying the size of the potential BSM
signals. This model-agnostic approach has its merits on many occasions, such as
analyses designed with minimal BSM assumptions. However, the ability to extract
the signal component is still very much demanded. It is possible to achieve
this goal, as suggested by the authors of ref.~\cite{GPR2017}, using a
stationary kernel, $S$, to model the signal component:

\[ S = A e^{-\frac{1}{2}(x_i - x_j)^2/{l_s}^2} e^{-\frac{1}{2}((x_i-m)^2 + (x_j - m)^2)/t^2}\]

where $A$ is a constant, $l_{s}$ refers to the length scale, $m$ specifies the
centre of the signal mass, and $t$ acts as the width of the
signal~\cite{GPR2017}. Therefore, the full GPR model becomes $S + \mathrm{C}(c)
\times \mathrm{RBF}(\ell)$. The pseudo-experiments done in
Section~\ref{sec:validation} using background-only spectra are used to
determine the bounds of the length scale of the background kernel. Most of
those pseudo-experiments report a best-fitted length scale between 0.6 and 0.7,
so $l_0$ ($l_1$) is set to be 0.6(0.7). Similarly, the hyperparameters
associated with the signal kernel are determined by fitting the signal
templates directly. While both $A$ and $l_{s}$ can float freely, $m$ and $t$ are only
allowed to change within a reasonable range of the fitted values. Given the
logarithmic transformation done to the dataset in Section~\ref{sec:datapre},
the signal in the original space is approximated by $e^{s_{i} + b_{i}} -
e^{b_{i}}$, where $s_{i}$ and $b_{i}$ are the signal and background components
in the $i$-th bin, predicted by GPR.  Figure~\ref{fig:extraction_examples}
shows the signal extraction results for the 800 GeV Gaussian-shaped signal with a
5\% width. The number of extracted signal events increases as the amount of
injected signal events, roughly following a linear response. However, the
extracted signal strength is systematically higher and has a wide spread. The
performance of signal extraction becomes very unstable for signal points close
to the \mjj boundaries. The data pre-processing, kernel selection and
hyperparameters all affect the signal extraction performance, which can be
further optimised.     

The sensitivity depends on the signal width, so the conclusions drawn in this
section are specific to Gaussian-shaped signals with a 5\% benchmark width, or
similar. The handling of wide signals has been a challenge in the functional
fit method, where the performance degrades as the signal width
increases~\cite{ATLASDijet}. The generality of GPR-based methods should be
tested in those more stringent cases in future works.  

\clearpage

\section{{\bfseries Robustness in HL-LHC}}
\label{sec:hllhc}

The authors of ref.~\cite{GPR2017} performed a test showing that the GPR-based background estimate has a stable $\chi^2$ result as the luminosity increases. Here, we extend this study to also include the KS and BH tests, in the HL-LHC scenario. The
analysis done in ref.~\cite{CMSMultib} uses 36.1 $\fb^{-1}$ of data, which
is only 1.2\% of the total integrated luminosity expected for HL-LHC. As a
consequence, events at the high \mjj tail expected in HL-LHC have not been
collected in this dataset. To obtain a test dataset that corresponds
to the HL-LHC scenario, the following procedure is applied: 

\begin{itemize}
\item Fit the 36.1 $\fb^{-1}$ \mjj spectrum obtained by GPR using $f(x) = p_{0}(1-x)^{p_1}x^{p_2}$, where $x = \mjj /6500$, starting from \mjj = 1000 GeV. $f(x)$ is the 3-parameter version of Function~\ref{equ:dijet}.
\item Use the fitted function to predict the yields at high mass tail that is not available in the 36.1 $\fb^{-1}$ dataset.    
\item Scale the whole spectrum, to the target integrated luminosity at the HL-LHC, which is 3000 $\fb^{-1}$.
\end{itemize} 

\begin{figure}[ht]
  \begin{center}
   \includegraphics[width=0.4\columnwidth]{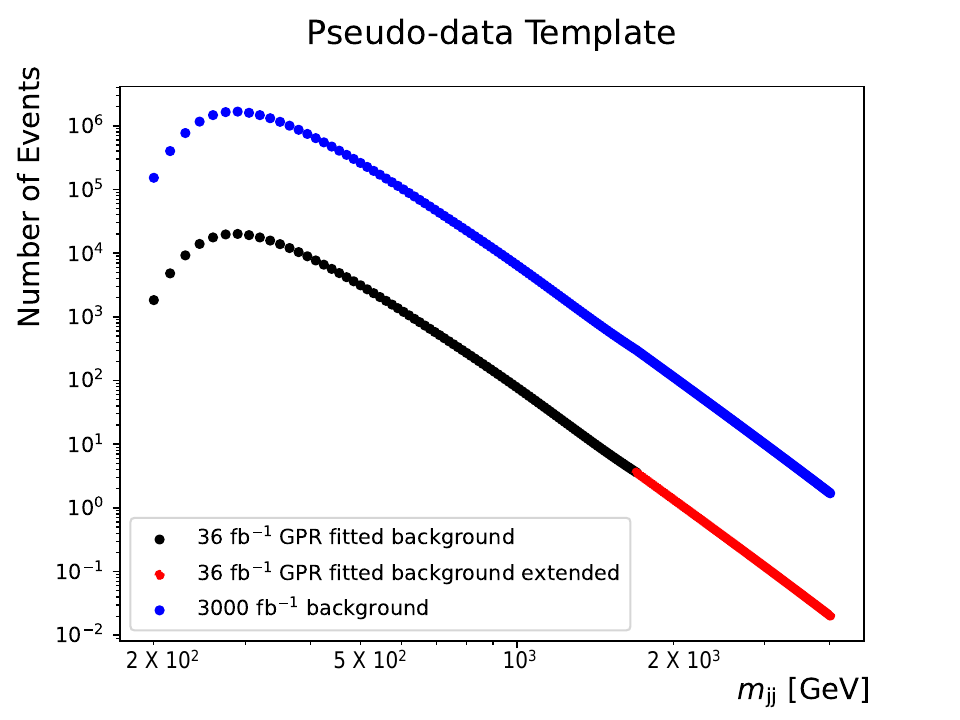}
   \includegraphics[width=0.4\columnwidth]{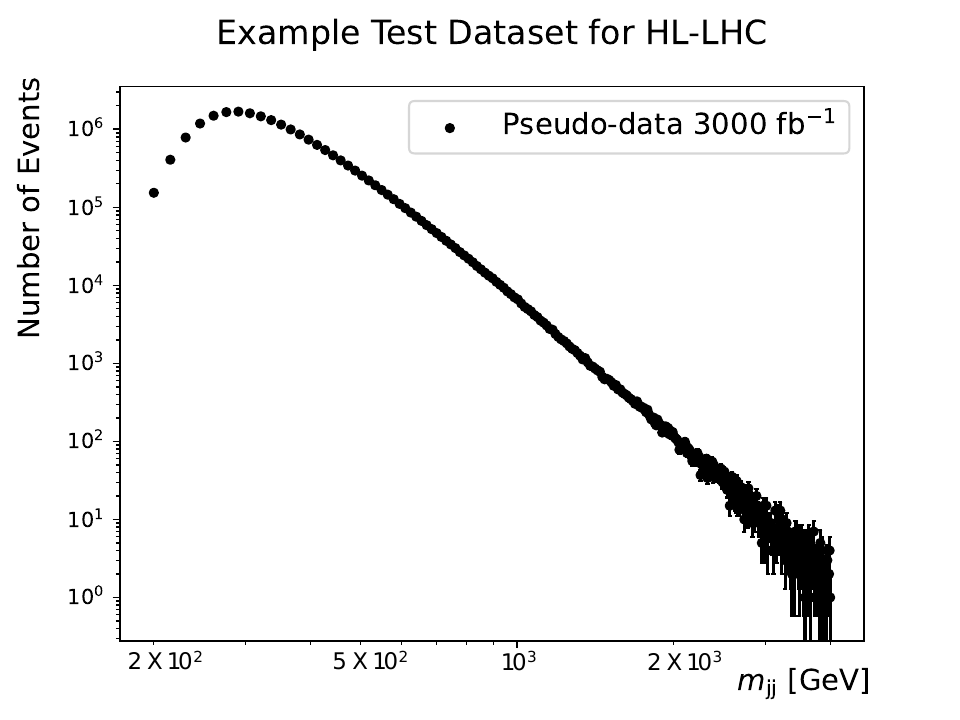}
  \caption{The smooth template used to generate pseudo-dataset (left) and one of the example pseudo-datasets (right), representing the HL-LHC scenario (3000 $\fb^{-1}$). \mjj refers to the invariant mass of the di-jet system.}
  \label{fig:bkg_fit_master_hllhc}
  \end{center}
\end{figure}

Figure~\ref{fig:bkg_fit_master_hllhc} shows the template used to generate the
pseudo-datasets for the HL-LHC scenario and one example dataset. It is acknowledged
that the events at the high mass tail may not accurately represent the real
data to be collected, but it suits the scope of this study, which is to
check the robustness of GPR against increasing luminosity.  

\begin{figure}[ht]
  \begin{center}
   \includegraphics[width=0.4\columnwidth]{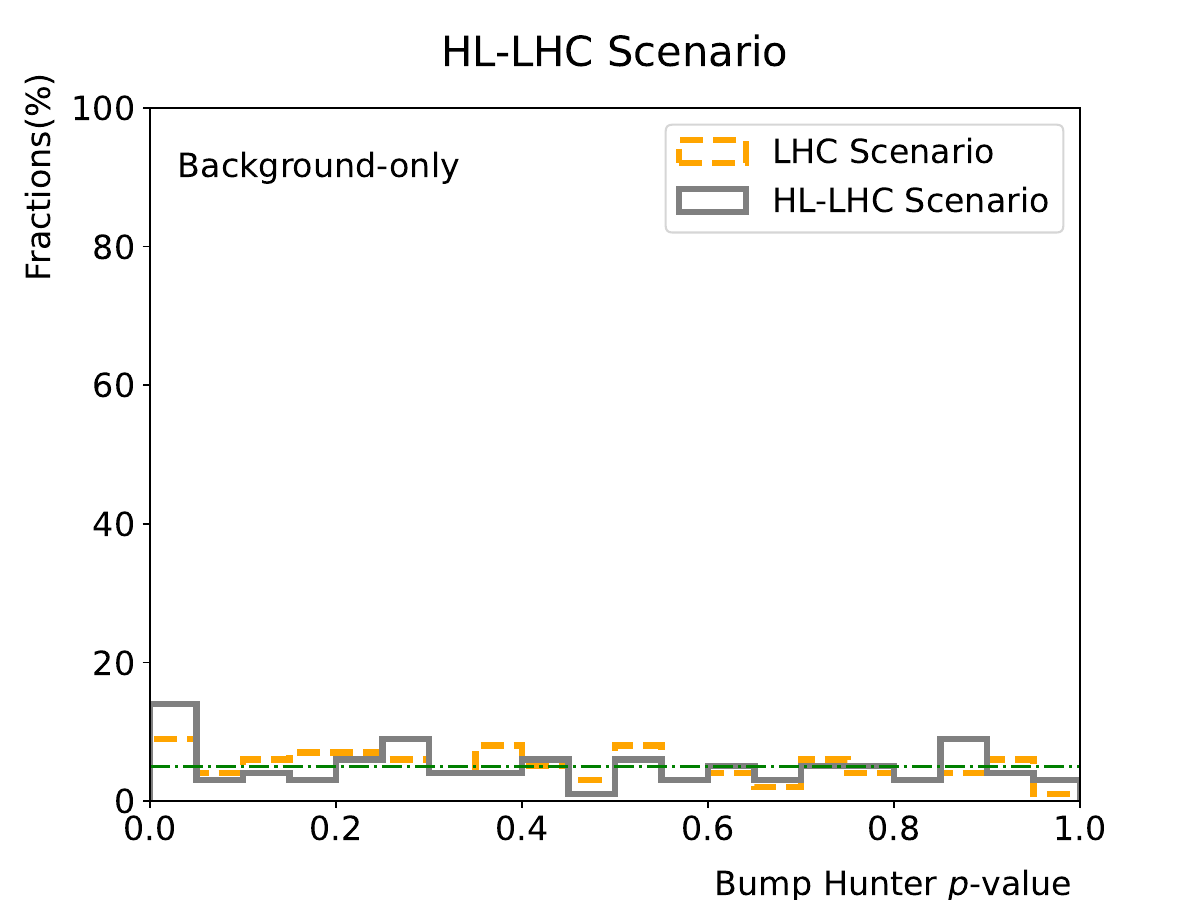}
   \includegraphics[width=0.4\columnwidth]{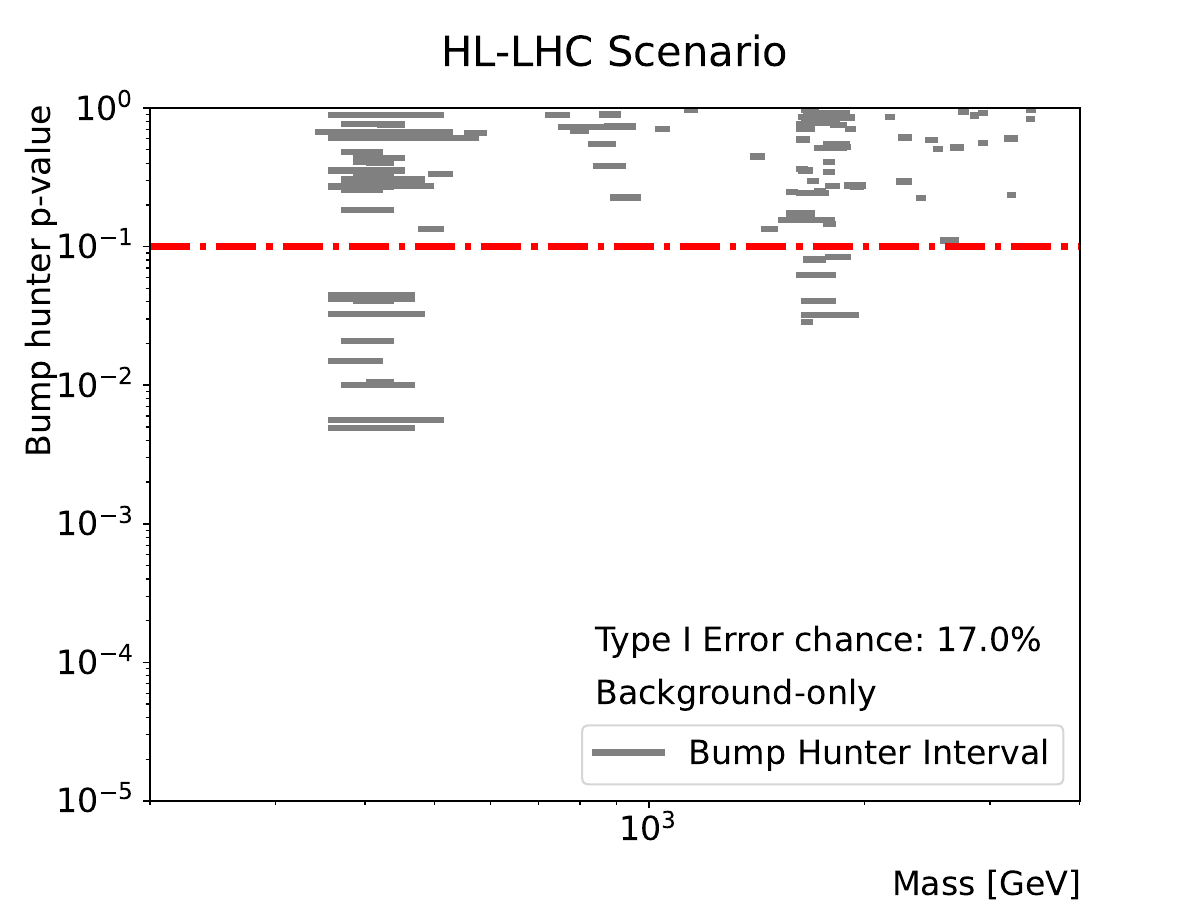}
  \caption{Left: BH $p$-value distributions of 100 background-only pseudo-experiments in the HL-LHC scenario (solid line). The green dashed line indicates the expected $p$-value distribution when no signal events are injected. Right: a summary of the BH $p$-values and the corresponding mass intervals (horizontal solid segments). Each solid horizontal segment comes from one pseudo-experiment. The horizontal dashed line corresponds to a critical value of 0.1. Flagged intervals with BH $p$-values above this threshold are considered not significant.}
  \label{fig:bh_bkg_only_hllhc}
  \end{center}
\end{figure}

The same set of tests are performed using the above pseudo-datasets. Since the
mass region is enlarged by a factor of 2.5 compared to the LHC test dataset,
the widest window considered in BH is increased from 10 to 20~\cite{BH,pyBumpHunter}. We only observe weak biases in the BH $p$-value test,
as shown in Figure~\ref{fig:bh_bkg_only_hllhc}. The results from KS and
$\chi^2$ tests are also very similar. 

\begin{figure}[ht]
  \begin{center}
   \includegraphics[width=0.35\columnwidth]{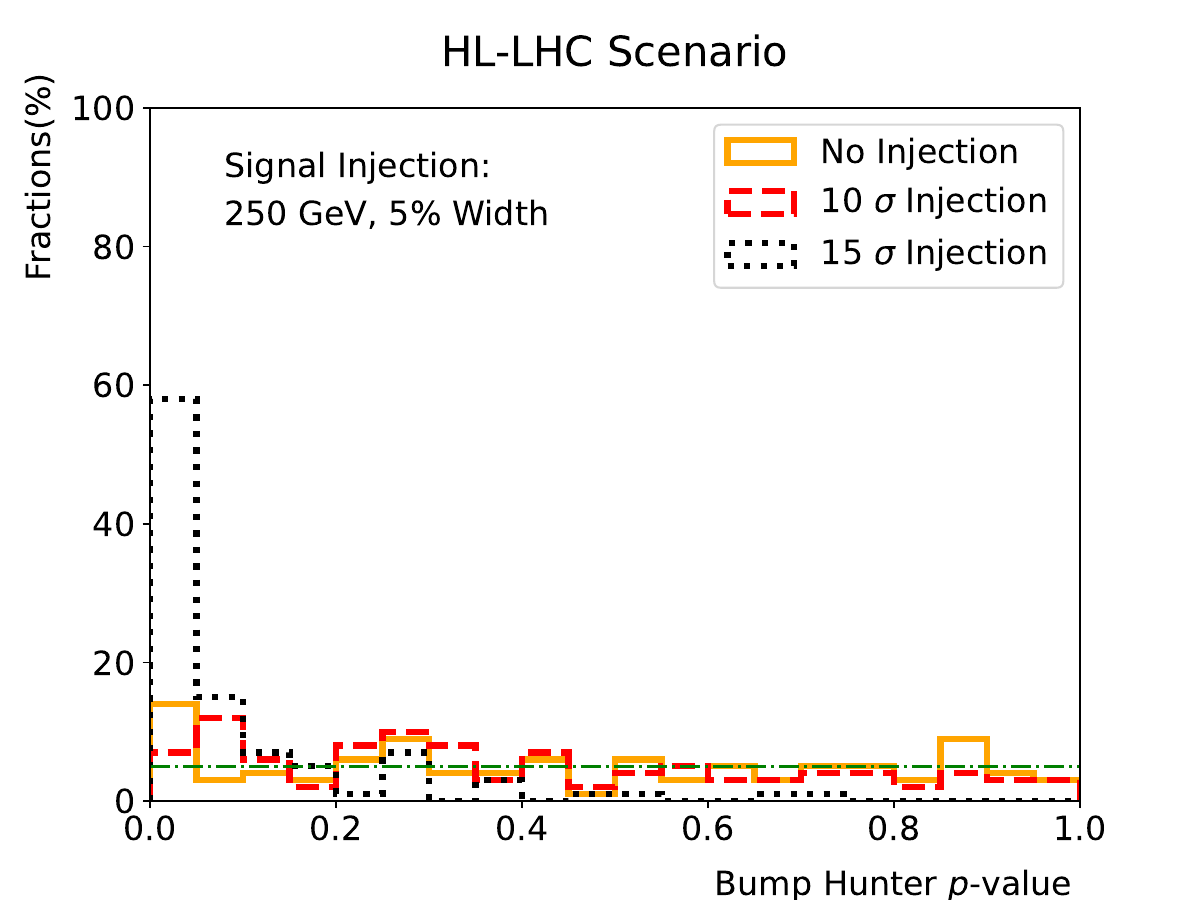}
   \includegraphics[width=0.35\columnwidth]{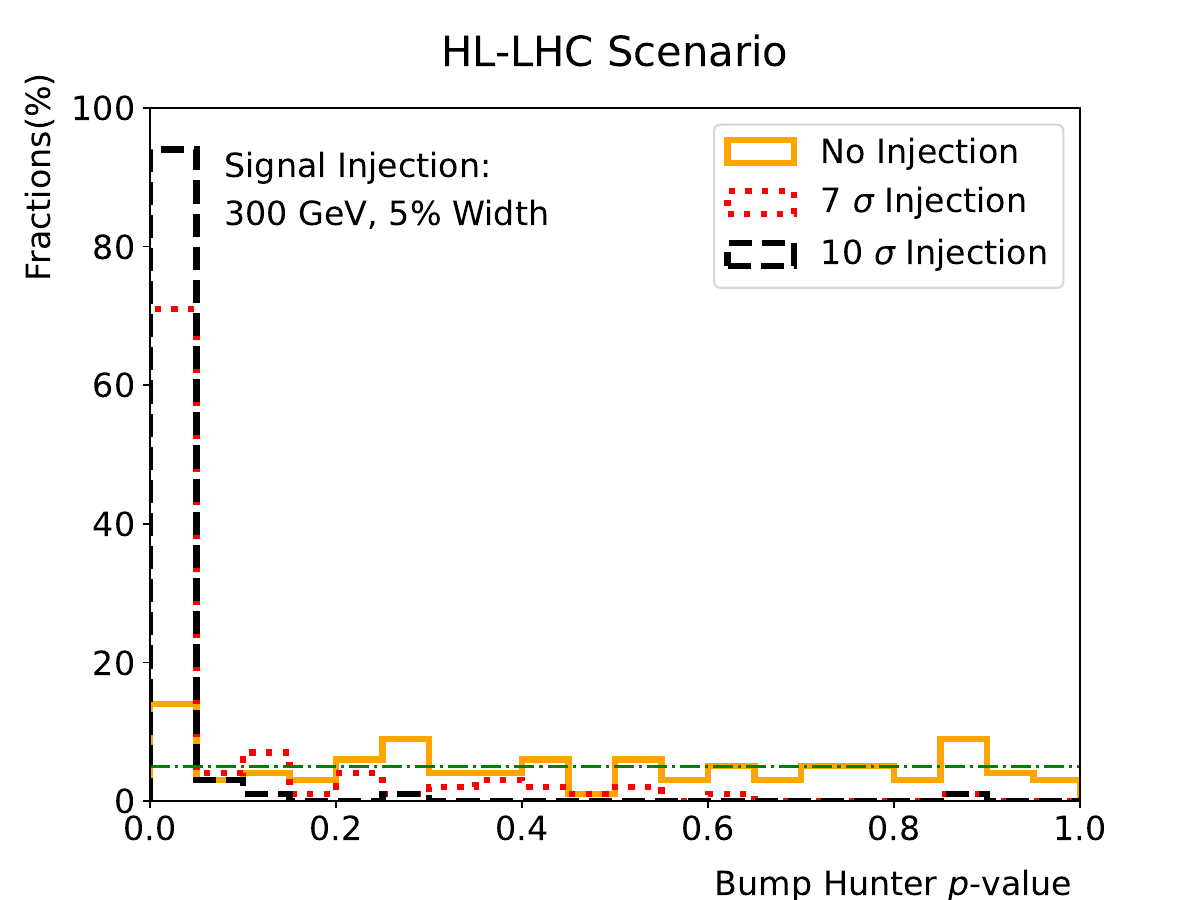}
   \includegraphics[width=0.35\columnwidth]{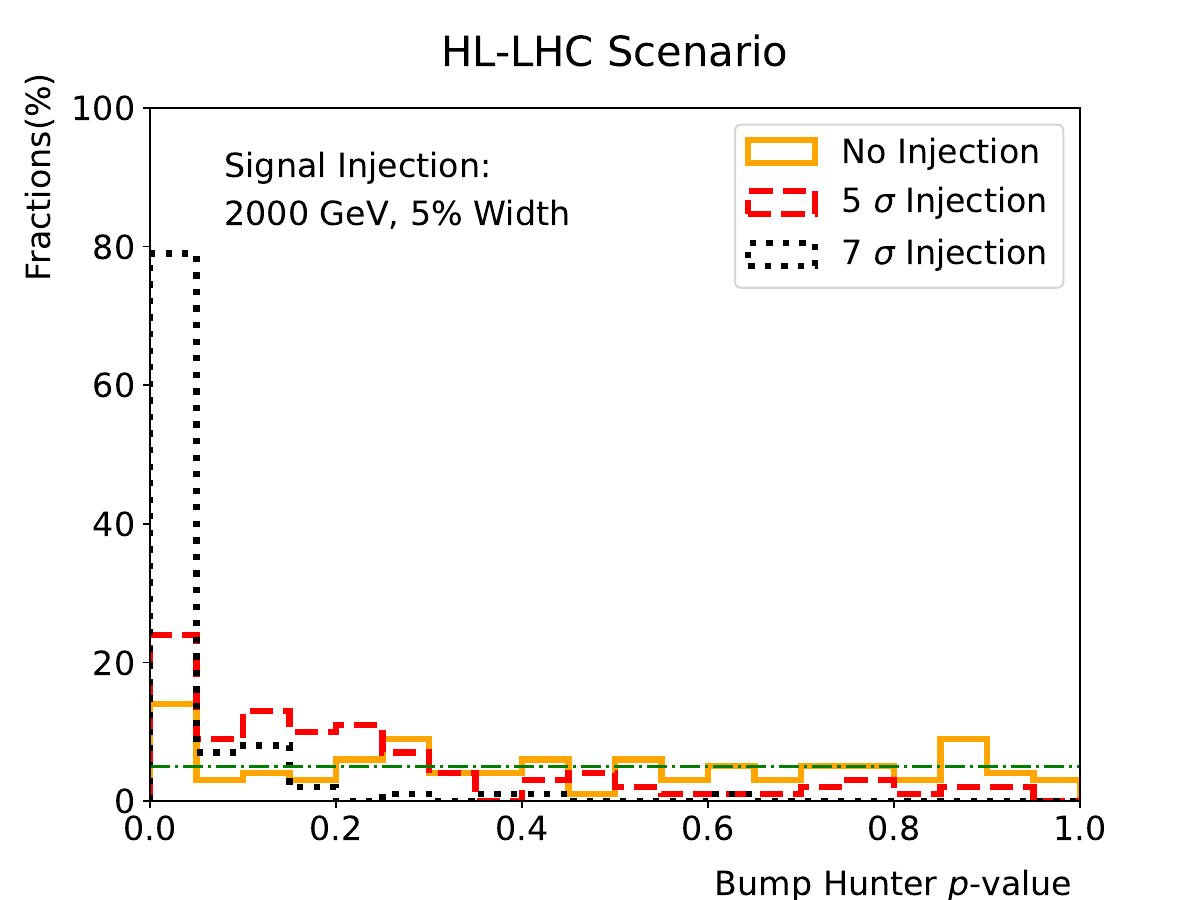}
   \includegraphics[width=0.35\columnwidth]{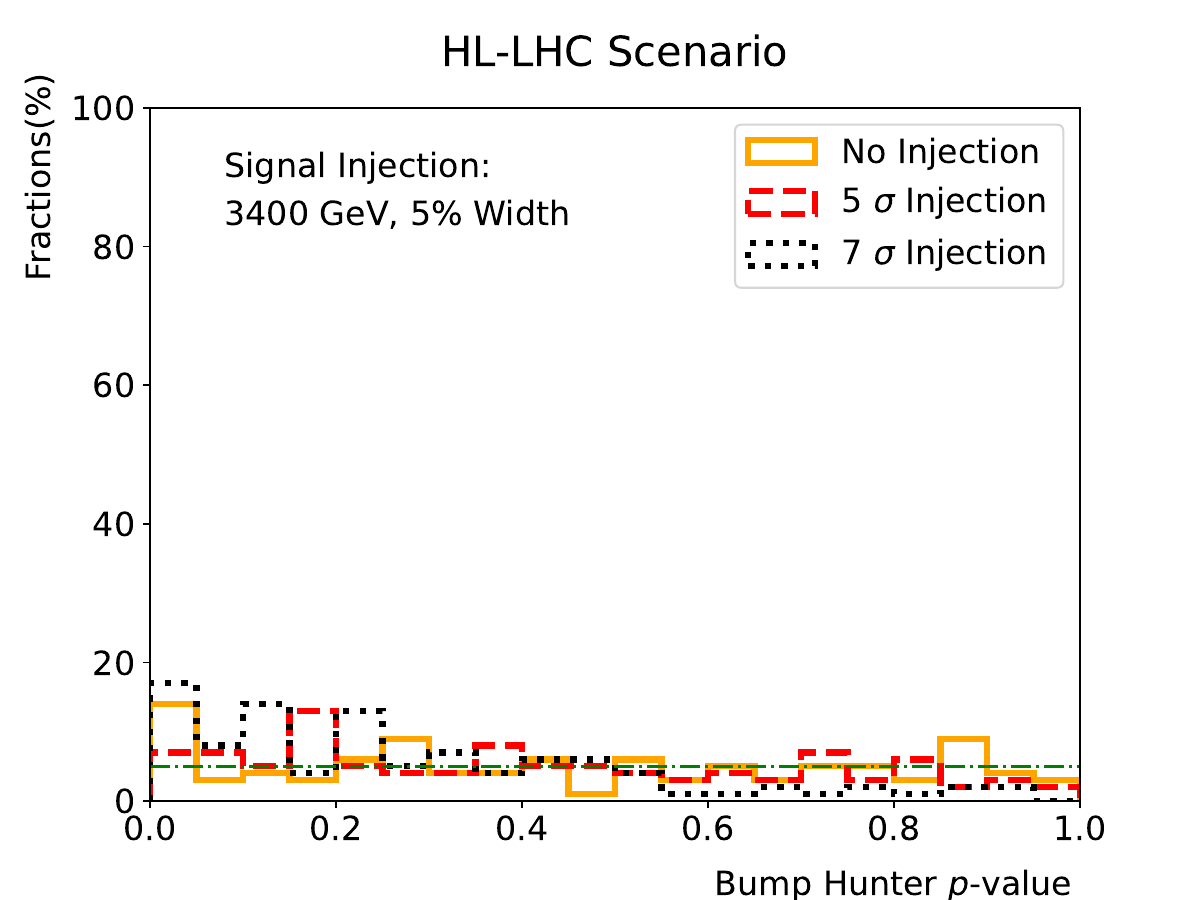}
  \caption{Summary of BH $p$-values obtained in the signal injection tests in the HL-LHC scenario, for the 250 GeV (upper-left), 300 GeV (upper-right), 2000 GeV (lower-left) and 3400 GeV (lower-right) signal mass points. The green dashed line indicates the expected $p$-value distribution when no signal events are injected.}
  \label{fig:injection_p_values_hl_lhc}
  \end{center}
\end{figure}

Since the mass region is extended up to 4 TeV given the expected HL-LHC
luminosity, the 800 GeV and 1400 GeV signal mass points are changed to 2000 GeV
and 3400 GeV, respectively, for the signal injection tests. The resulting BH
$p$-values are shown in Figure~\ref{fig:injection_p_values_hl_lhc}, which are
similar to the LHC scenario. 

\begin{figure}[ht]
  \begin{center}
   \includegraphics[width=0.35\columnwidth]{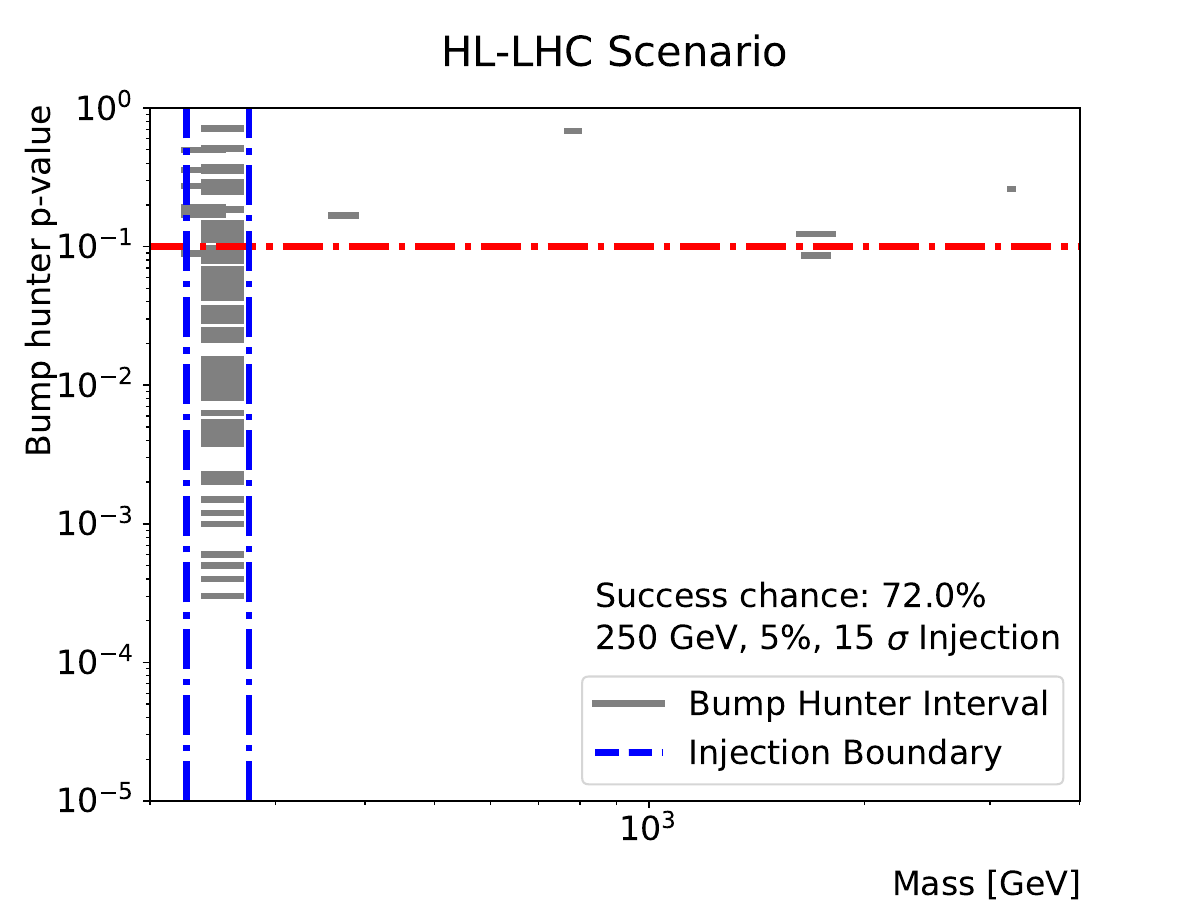}
   \includegraphics[width=0.35\columnwidth]{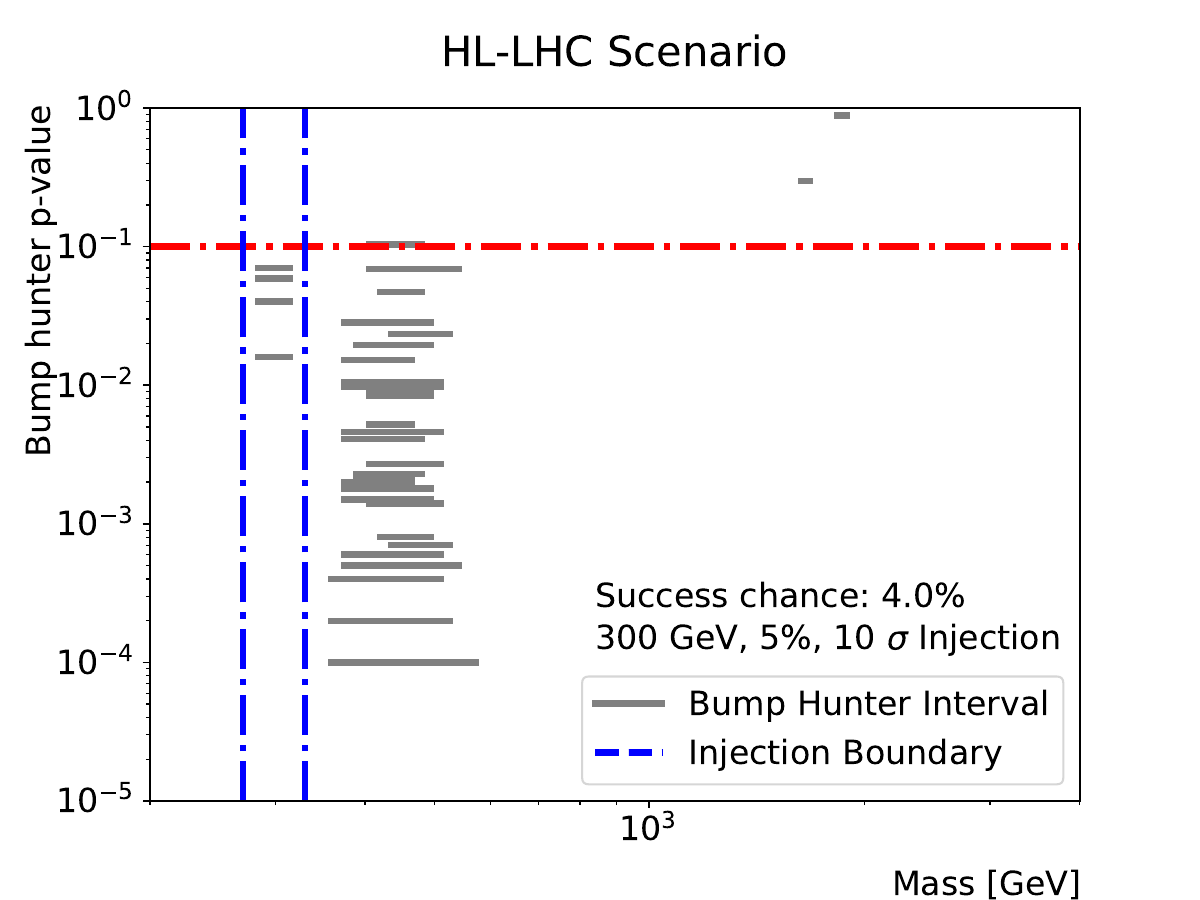}
   \includegraphics[width=0.35\columnwidth]{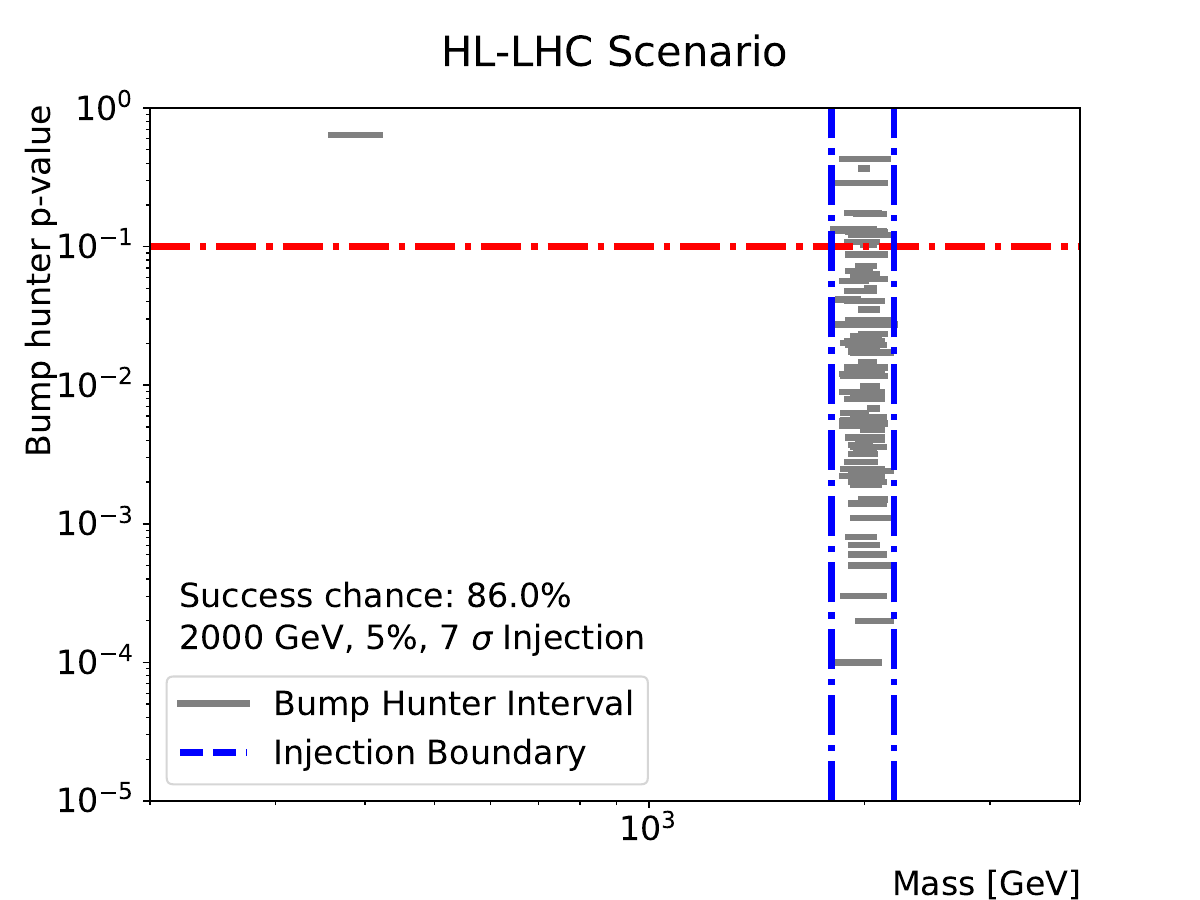}
   \includegraphics[width=0.35\columnwidth]{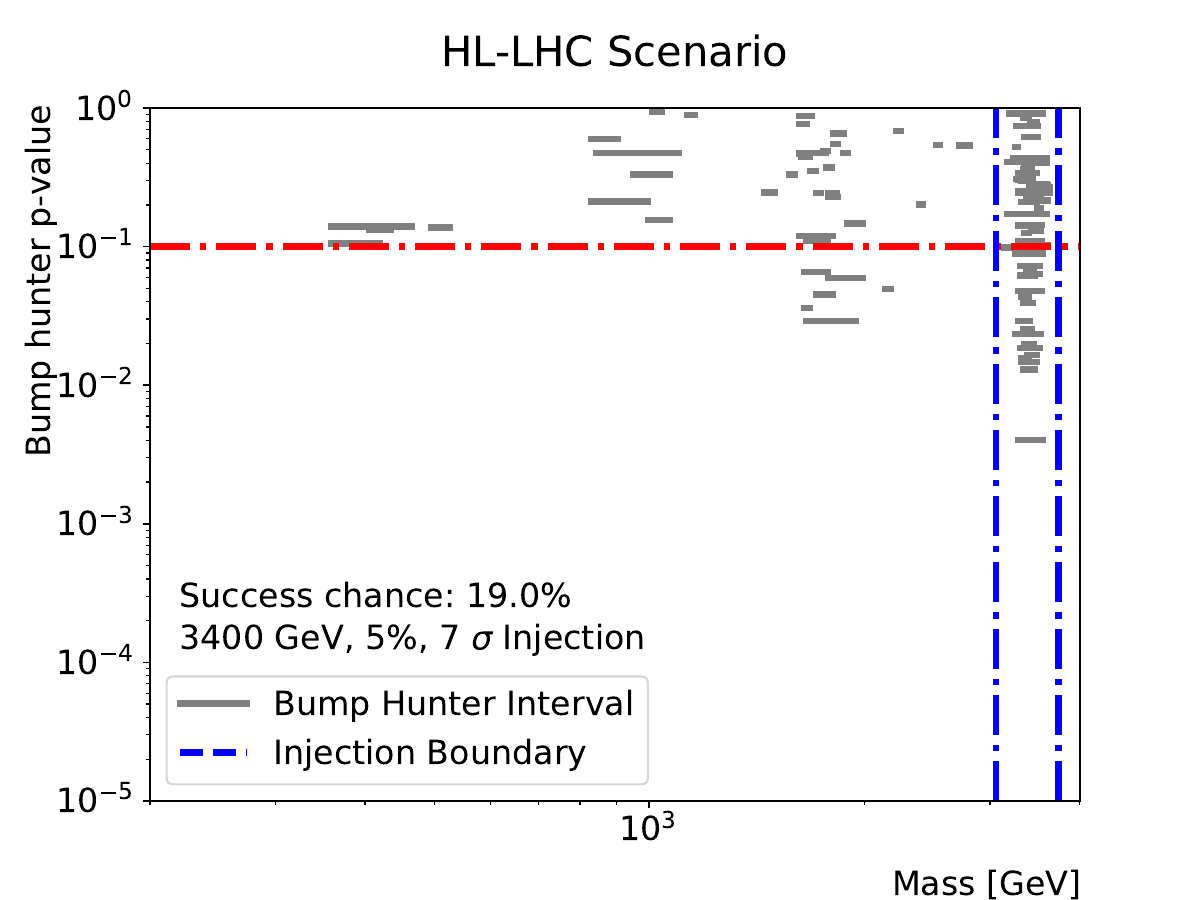}
  \caption{Summary of the BH $p$-values and the corresponding mass intervals (solid horizontal segment), for the 250 GeV (upper-left), 300 GeV (upper-right), 2000 GeV (lower-left) and 3400 GeV (lower-right) signal mass points, in the HL-LHC scenario. Each solid horizontal segment comes from one pseudo-experiment. The horizontal dashed line corresponds to a critical value of 0.1. Flagged intervals with BH $p$-values above this threshold are considered not significant. The vertical dashed-dotted lines represent the \mjj region where signal events are injected.}
  \label{fig:injection_p_values_edges_hl_lhc}
  \end{center}
\end{figure}

\begin{figure}[ht]
  \begin{center}
   \includegraphics[width=0.35\columnwidth]{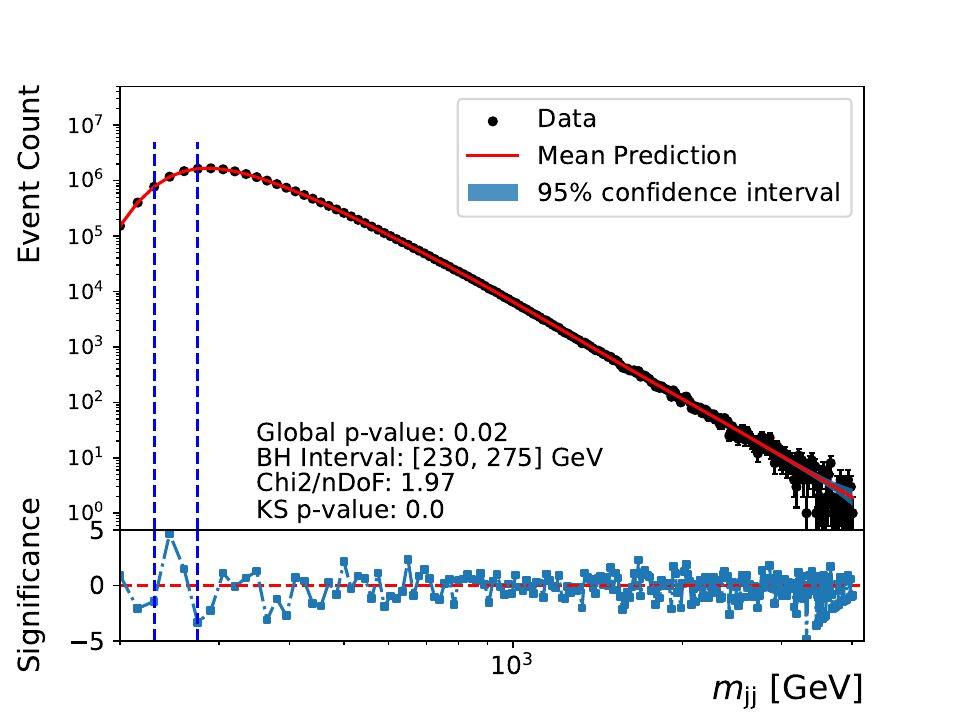}
   \includegraphics[width=0.35\columnwidth]{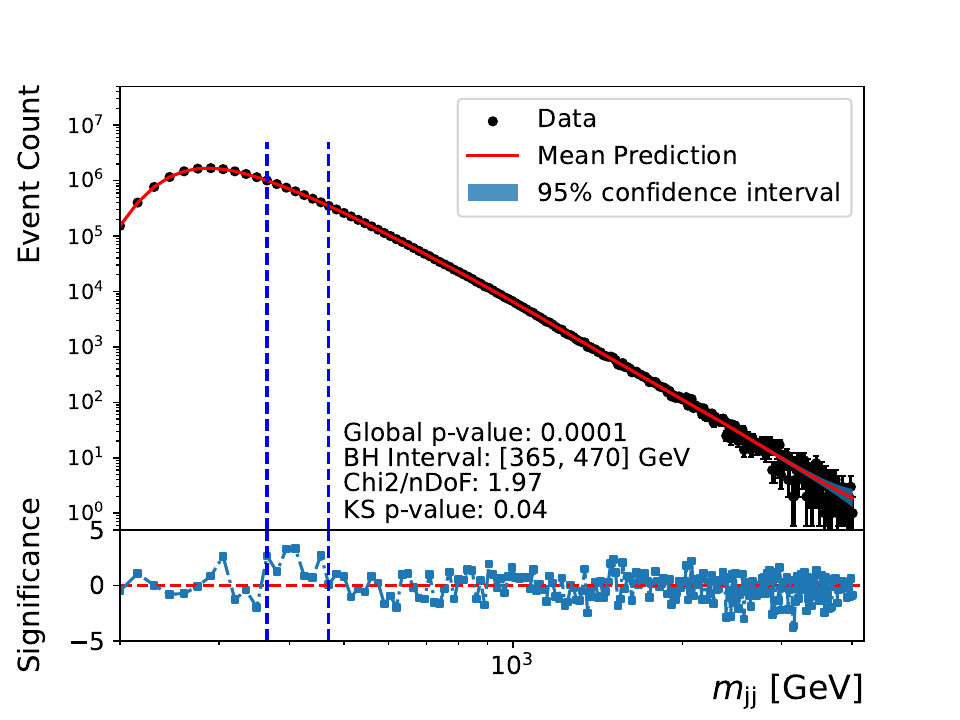}
   \includegraphics[width=0.35\columnwidth]{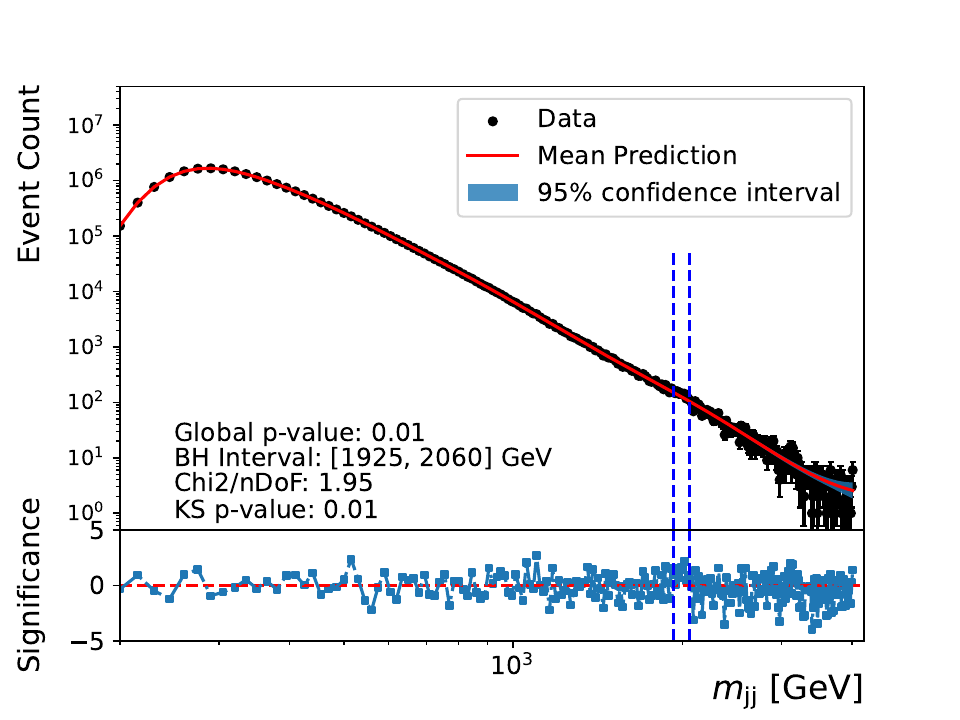}
   \includegraphics[width=0.35\columnwidth]{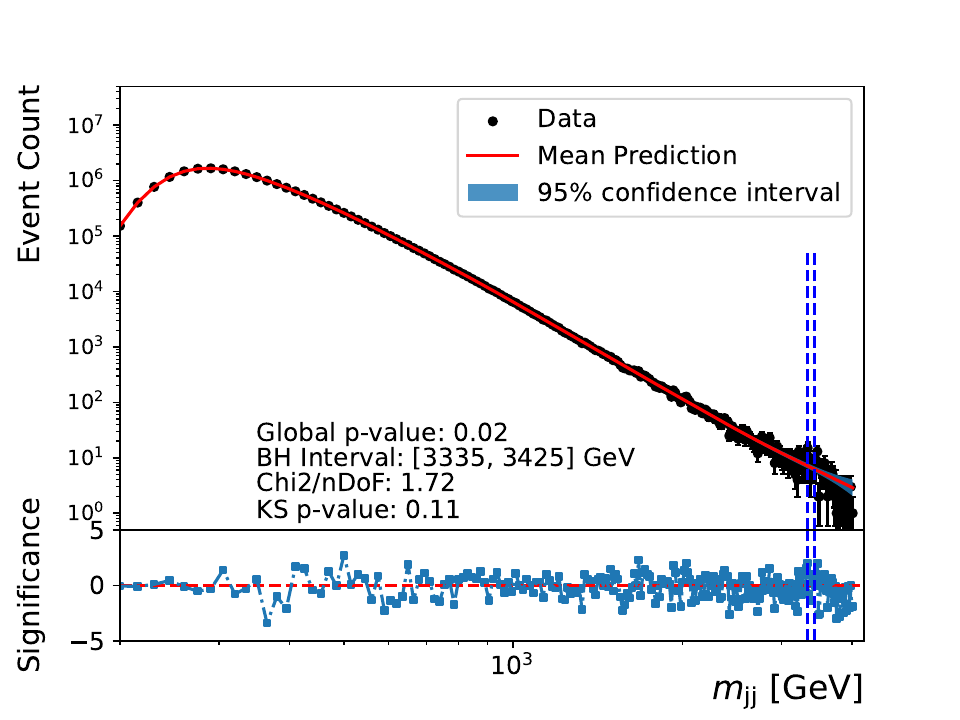}
  \caption{Comparison between the signal-injected pseudo-data (solid point) and the background estimate from GPR (solid line), for the 250 GeV (upper-left), 300 GeV (upper-right), 2000 GeV (lower-left) and 3400 GeV (lower-right) signal mass points, in the HL-LHC scenario. The vertical dashed lines indicate the boundaries of the most significant deviation reported by BH. The bottom panels present the significance calculated for each mass bin.}
  \label{fig:injection_examples_hllhc}
  \end{center}
\end{figure}

However, BH fails to report the most significant deviations at the correct
location when the 300 GeV signal is injected, although the fraction of BH
$p$-values below 0.1 is high, as shown in
Figure~\ref{fig:injection_p_values_edges_hl_lhc}. It is already observed in the
LHC scenario, but the impact is enhanced due to a much larger luminosity.
Figure~\ref{fig:injection_examples_hllhc} shows random examples of the
pseudo-experiments that have successfully reported a $p$-value below 0.1, with
the BH intervals marked as well.

The same signal extraction procedure is tested for the HL-LHC scenario as well,
using the 2000 GeV Gaussian-shaped signal with a 5\% width. The performance is
similar to that of the LHC scenario. The signal extraction is systematically
higher, with a wide spread.

\begin{figure}[ht]
  \begin{center}
    \includegraphics[width=0.45\columnwidth]{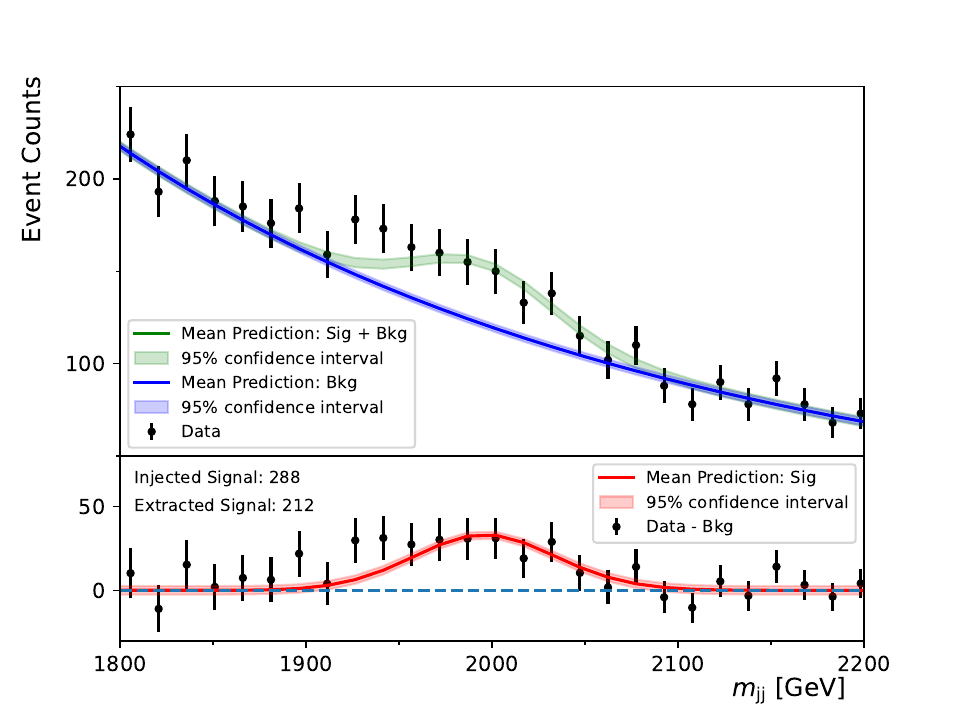}
    \includegraphics[width=0.45\columnwidth]{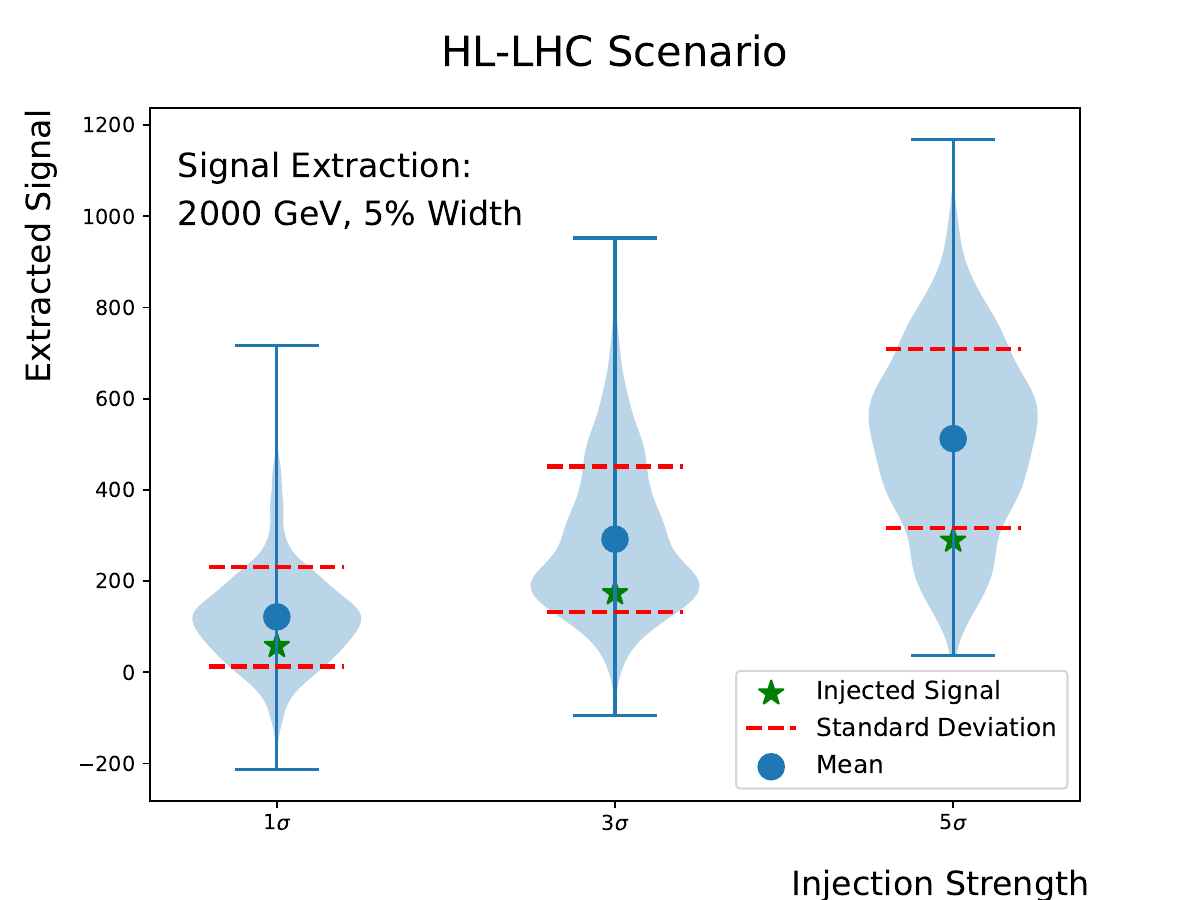}
    \caption{An example of signal extraction test done for the 2000 GeV Gaussian-shaped signal with a 5\% width (left), and the summary of test results with different injected signal strengths (right), in the HL-LHC scenario. The numbers of injected signal events are indicated by the green stars, while the blue dots represent the means of extracted signal. The width of the shaded bands corresponds to the density of a given number of extracted signal events. The red dotted lines are the one standard deviation boundaries.}
    \label{fig:extraction_examples_hllhc}
    \end{center}
\end{figure}

The overall performance of the same GPR model decreases slightly in the HL-LHC
scenario if the hyperparameters are not re-optimised (HL-LHC results using
re-optimised hyperparameters are discussed in
Appendix~\ref{app:hllhc_tuning}). It is remarkable as the luminosity is
increased by two orders of magnitude. The hyperparameters permit the users to
tune the model with great flexibility, while the traditional functional fit
method limits the users to a given function family. Due to the sizeable
correlations between the free parameters, the fit performance of a function is
saturated after reaching a certain number of free parameters. In contrast, the
GPR-based background modelling technique can offer a solution that is valid
through the entire lifetime of the LHC. 

\clearpage     

\section{{\bfseries Summary}} 
\label{sec:summary}

In this work, the performance of a GPR-based background estimation method is
thoroughly investigated, with a set of tests performed using a background
spectrum reported in a CMS search~\cite{CMSMultib}. The background is
challenging as it also includes the region before the smoothly falling part.
The CMS search divides the background into three separate regions, with each fitted
by a different function. Owing to the flexible nature of GPR, this
delicate spectrum can be handled in a much simpler way. We point out that one
can rely on the most commonly adopted RBF kernel to achieve excellent
performance if the $L_2$ regularisation is tuned accordingly. Thus, for a
GPR-based background estimate model, the set of hyperparameters includes the
bounds of the kernel parameters and the regularisation matrix. The discovery
potential is evaluated using the \textsc{BumpHunter} algorithm~\cite{BH}. With
a minimally optimised set of hyperparameters, we observe promising sensitivity
to hypothetical narrow resonances. In addition, the background is projected to
the HL-LHC to test how robust GPR is against increasing luminosity. Unlike
traditional functional fit methods that have already been challenged seriously
during Run 2, the performance of the GPR-based strategy is
remarkably stable even for the HL-LHC scenario. The signal extraction procedure
is tested as well, without optimisation for this specific task. While
the outcome is positive, it is obvious that the performance should be
improved. An automatic and systematic way to tune a GPR-based model to suit all
major use cases is of high value, which is an interesting topic for future
studies.  

Recent LHC analyses have started experimenting with GPR on various
occasions~\cite{CMSTLA,ATLASHC,ATLASyy}, but functional fit is still the most
widely embraced method in resonance searches. It is in part due to its long
historical success, and there are well-established procedures to address topics
such as systematic uncertainties and to integrate it into the statistical
analysis~\cite{ATLASBkgNote}. In this work, we illustrate that GPR can satisfy
the common standards imposed for searches using the \textsc{BumpHunter}
algorithm, even if a very difficult background shape is under consideration. It
is clear that GPR is already suitable for certain types of searches at the LHC,
and we look forward to more results adopting this method. Its robustness
against increasing luminosity allows the community to develop background
estimate strategies that are valid throughout the entire lifetime of the
(HL)-LHC, greatly simplifying the workflows. It is not discussed here how to evaluate the major systematic
uncertainties associated with GPR and how the hyperparameters can impact them.
We leave this topic for future works. 

\section*{{Acknowledgments}}
GPR-based methods have been investigated by the ATLAS collaboration for several
years. We have learned a lot from numerous talks given by our colleagues,
and those pioneering GPR applications in ATLAS analyses. We thank Rachel
Hyneman for helping with the signal extraction workflow, and Marco Montella for
valuable suggestions. Jackson B. is supported by the STFC UCL Centre for Doctoral Training in Data Intensive Science (grant ST/P006736/1), including by departmental and industry contributions. B.X. Liu is supported by Shenzhen Campus of the Sun
Yat-sen University under project 74140-12240013. B.X. Liu appreciates the
support from Guangdong Provincial Key Laboratory of Gamma-Gamma Collider and
Its Comprehensive Applications, and the support from Guangdong Provincial Key
Laboratory of Advanced Particle Detection Technology.  

\appendix
\section{Signal injected spectra}
\label{app:sig_inj}
\begin{figure}[ht]
  \begin{center}
    \includegraphics[width=0.4\columnwidth]{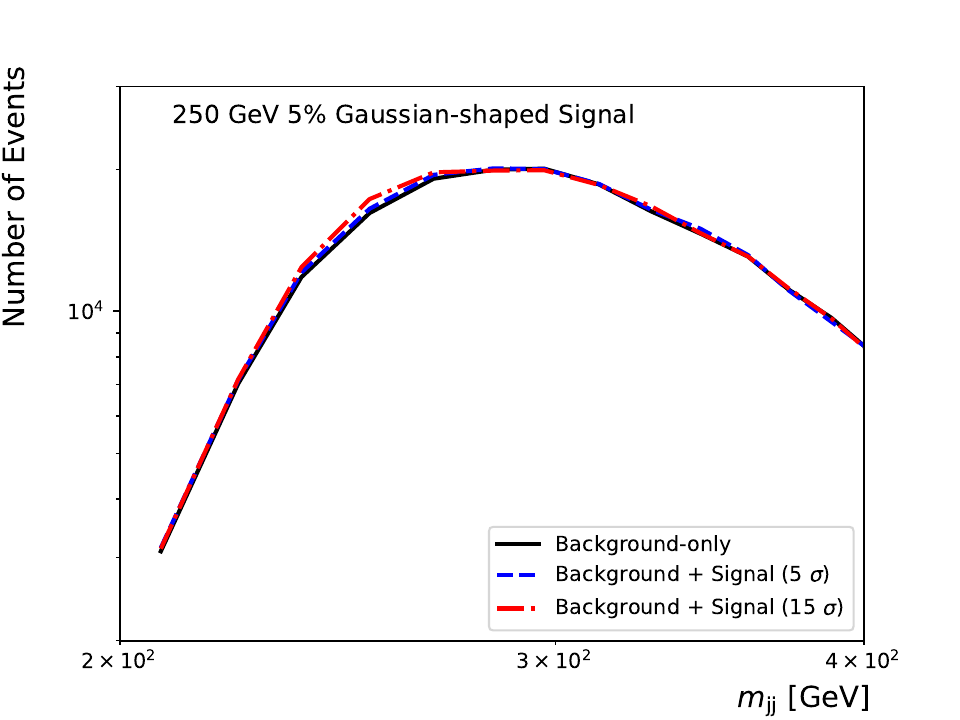}
    \includegraphics[width=0.4\columnwidth]{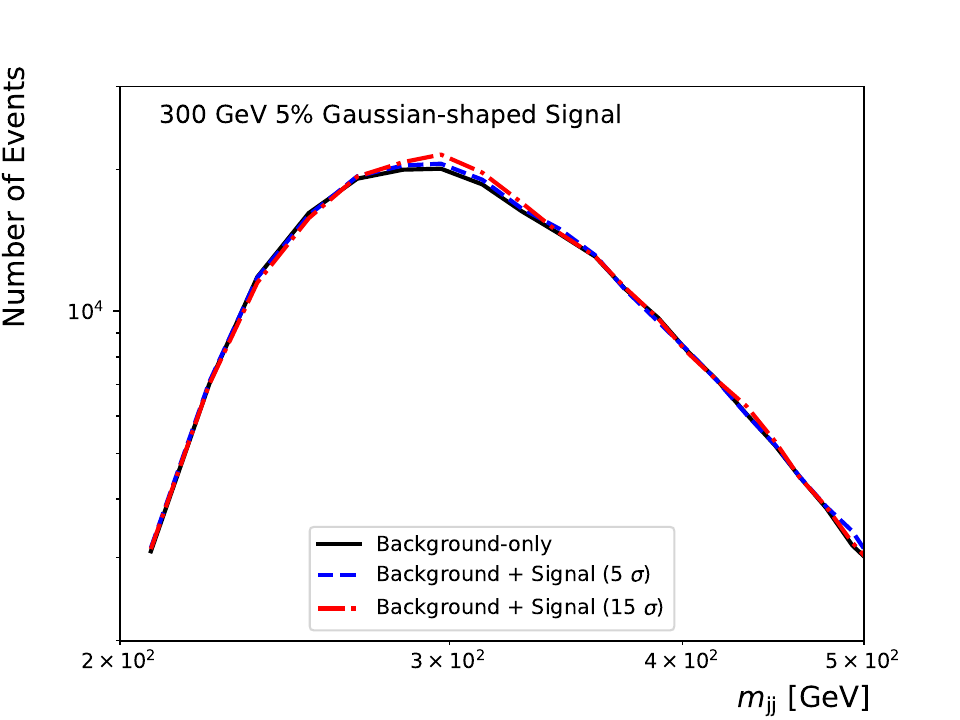}
    \includegraphics[width=0.4\columnwidth]{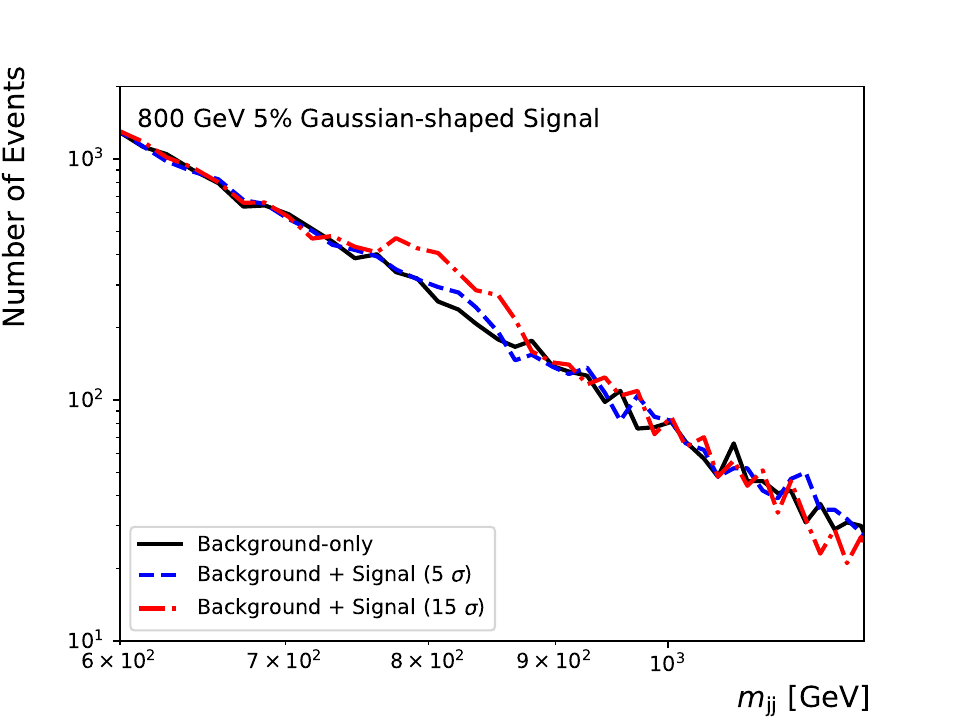}
    \includegraphics[width=0.4\columnwidth]{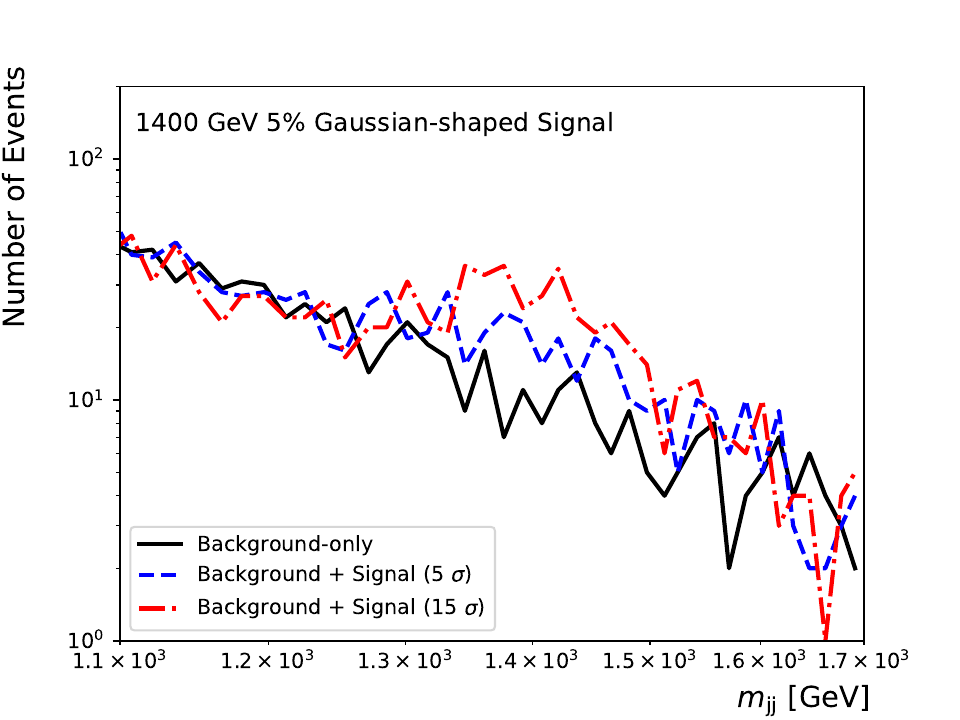}
    \caption{Examples of signal injected spectra for the 250, 300, 800 and 1400 GeV Gaussian-shaped signals with a 5\% width. Both the background-only (solid black line), the 5-$\sigma$ injection (blue dashed line) and the 15-$\sigma$ injection (red dotted-dashed line) spectra are shown. The mass range is zoomed in to better visualise the region near the signal.}
    \label{fig:sig_inj_spec}
  \end{center}
\end{figure}

The intrinsic difficulty to retain good sensitivity in the region below the
smoothly falling part of the spectrum can be appreciated with examples such as
Figure~\ref{fig:sig_inj_spec}. When injecting a 5\% Gaussian-shaped signal with
a mass of 250 GeV, the injected spectrum has no clear localised structures. It
is because this mass point is very close to the starting point of the fit, and
the spectrum has a much steeper slope. Similarly, if a 300 GeV signal is
considered, which is right at the plateau of the background, no obvious
localised structures present neither. Those cases challenge the very assumption
of the search strategy that the presence of a significant signal will create
bumps. The ideal scenario is the 800 GeV signal, where the background on both
sides follow the same trend. As the signal mass gradually approaches the end of
the background distribution, the fit starts lacking constraints on the high
mass side and suffering from large statistical fluctuations. 

The above statements apply to both the functional fit method and the GPR-based
method. Despite their entirely different underlying methodologies, they
both rely solely on data. If the signal-injected background spectrum is as
smooth as the background-only one, neither method ought to achieve optimal
sensitivity.     

\appendix
\section{Impacts of the $L_2$ regularisation}
\label{app:hllhc_tuning}
In the HL-LHC scenario, it is observed that the background modelling biases are
increased, as shown in Figure~\ref{fig:bh_bkg_only_hllhc}. Consequently, a large
fraction of most significant BH-intervals being reported at the wrong locations
when the 300 GeV signal events are injected, as reported by
Figure~\ref{fig:injection_p_values_hl_lhc}. As mentioned in
Section~\ref{sec:reg}, the multiplication factor $f_{i}$ can serve as a set of
hyperparameters. In this appendix, we demonstrate that tuning $f_{i}$ helps
with achieving better background modelling performance for the HL-LHC scenario. 

\begin{figure}[ht]
  \begin{center}
    \includegraphics[width=0.4\columnwidth]{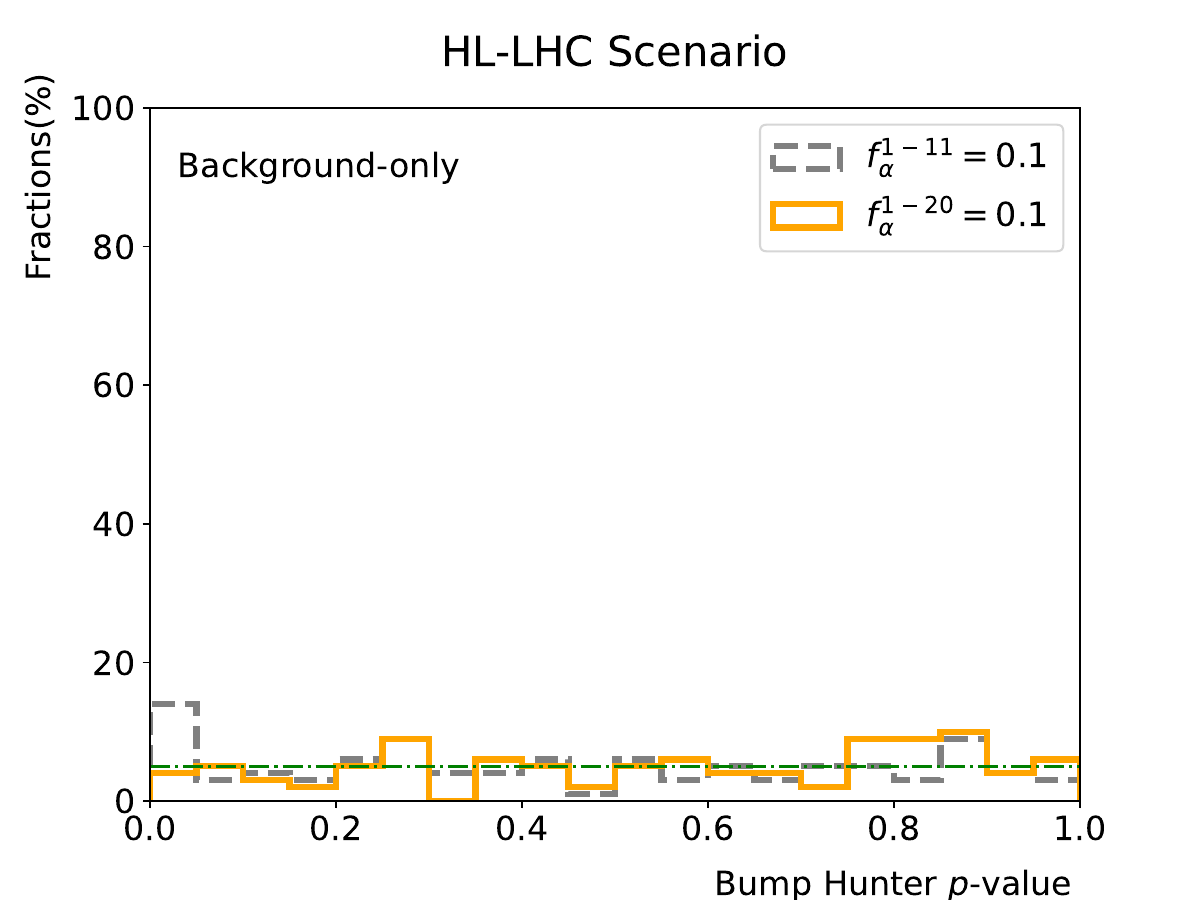}
    \includegraphics[width=0.4\columnwidth]{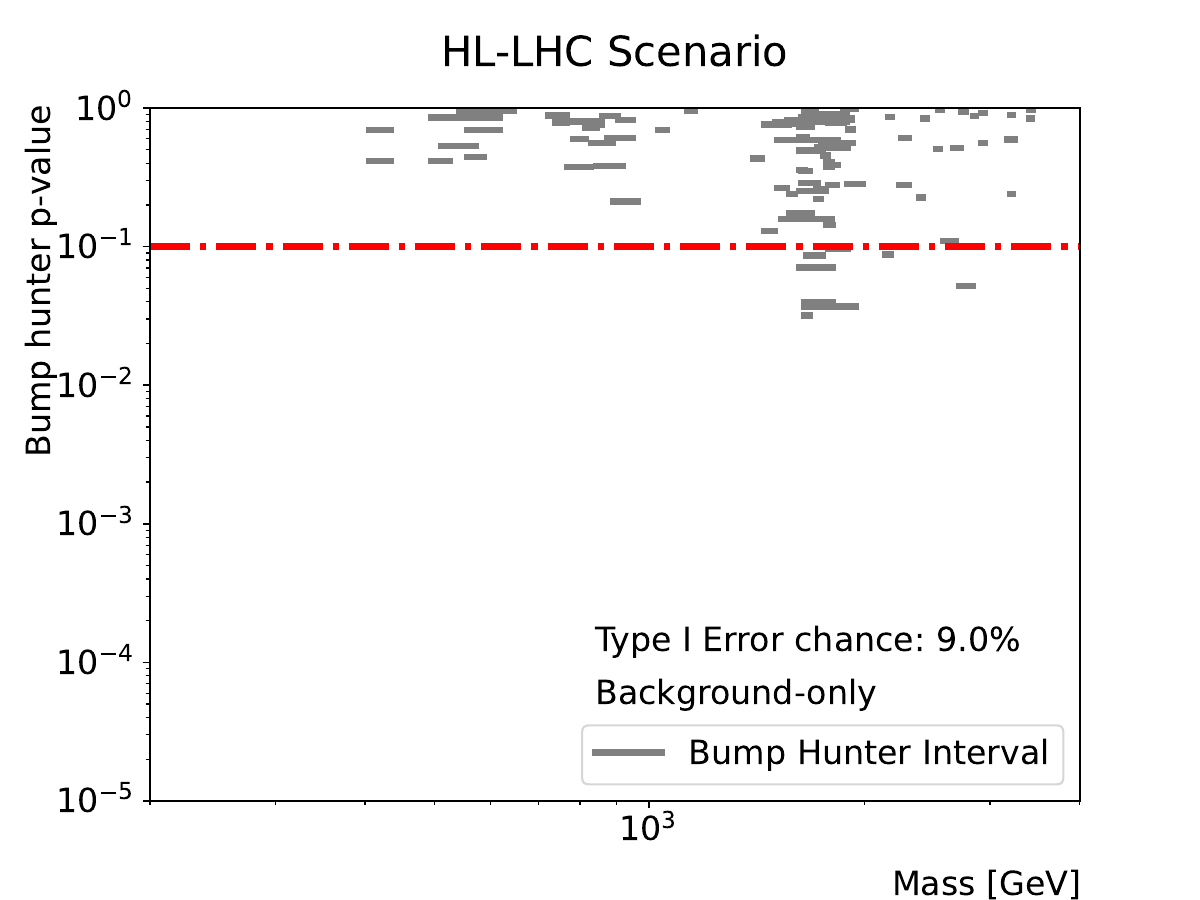}
    \caption{Left: BH $p$-value comparison between results using original tuned hyperparameters ($f_{1-11} = 0.1$) and the ones using re-tuned hyperparameters ($f_{1-20} = 0.1$), for the HL-LHC scenario. The green dashed line indicates the expected p-value distribution when no signal events are injected. Right: summary of the BH $p$-values and the corresponding mass intervals (solid horizontal segment), for the background-only tests, in the HL-LHC scenario. Each solid horizontal segment comes from one pseudo-experiment. The horizontal dashed line corresponds to a critical value of 0.1. Flagged intervals with BH $p$-values above this threshold are considered not significant.}
    \label{fig:hllhc_alpha_comparison_bkg_only}
  \end{center}
\end{figure}

\begin{figure}[ht]
  \begin{center}
    \includegraphics[width=0.4\columnwidth]{plots/bumphunter_p-value_edges_bkg_sig_300_0p5_10_hllhc.pdf}
    \includegraphics[width=0.4\columnwidth]{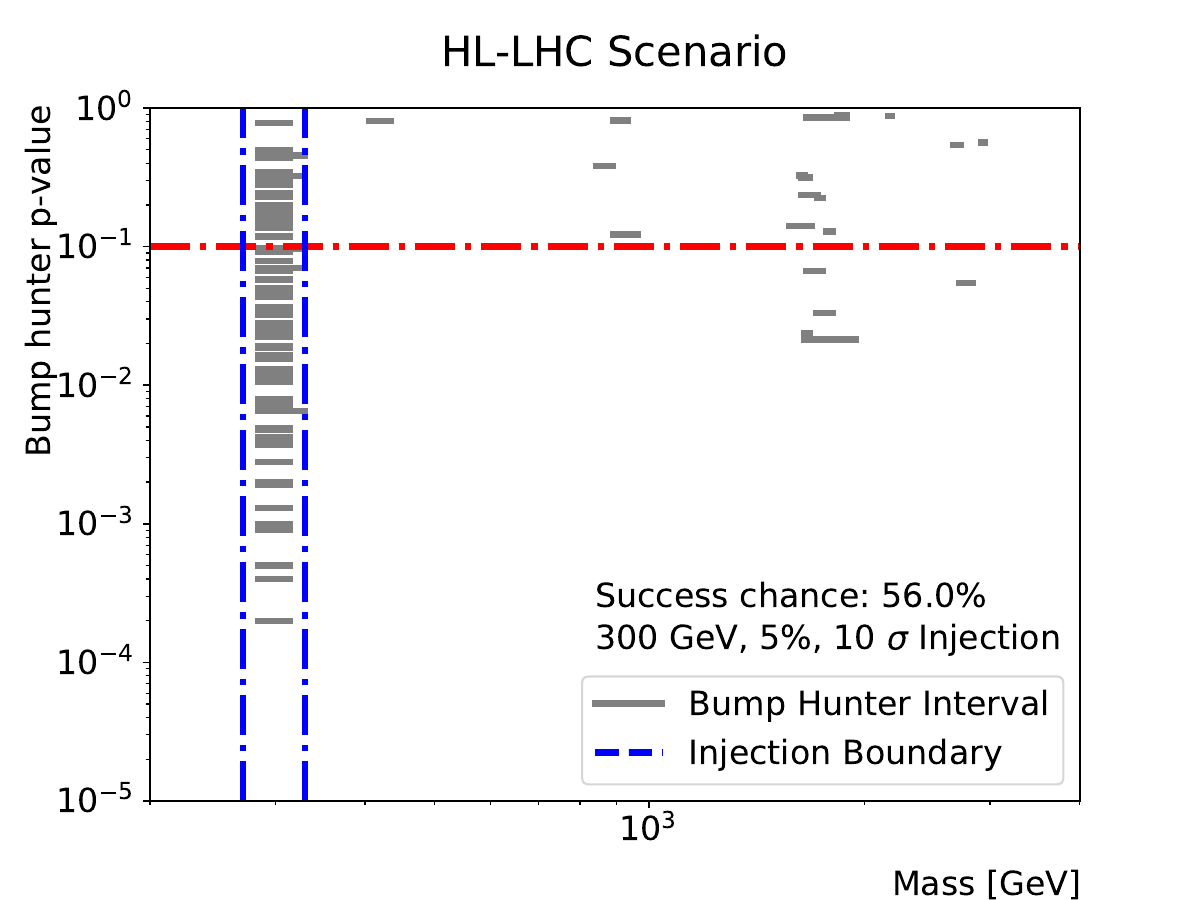}
    \caption{Comparisons of the BH $p$-values and the corresponding mass intervals between results using original tuned hyperparameters ($f_{1-11} = 0.1$) and the ones using re-tuned hyperparameters ($f_{1-20} = 0.1$), for the 300 GeV signal injection test, in the HL-LHC scenario. Each solid horizontal segment comes from one pseudo-experiment. The horizontal dashed line corresponds to a critical value of 0.1. Flagged intervals with BH $p$-values above this threshold are considered not significant. The vertical dashed-dotted lines represent the \mjj region where signal events are injected.}
    \label{fig:hllhc_alpha_comparison_injection}
  \end{center}
\end{figure}

Both Figure~\ref{fig:bh_bkg_only_hllhc} and
Figure~\ref{fig:injection_p_values_hl_lhc} indicate that the background
modelling is suboptimal around 400 GeV. Therefore, we modified the corresponding
$f_i$ to enhance the performance in this region. The original $f_{1-11} = 0.1$
is updated to $f_{1-20} = 0.1$.
Figure~\ref{fig:hllhc_alpha_comparison_bkg_only} compares the BH $p$-values
obtained in the background-only tests and the BH intervals reported using the
updated multiplication factor. The systematic pattern seen around 400 GeV before is
greatly mitigated, as expected.
Figure~\ref{fig:hllhc_alpha_comparison_injection} compares the 300 GeV signal
injection test results using the original ($f_{1-11} = 0.1$) and
re-tuned $f_{1-20} = 0.1$. The latter correctly reports
the excess at the location where the signal events are injected. 

This exercise demonstrates that the $L_2$ regularisation has a significant impact
on the background modelling performance. Tuning the corresponding
hyperparameters helps adapt the GPR model to various new conditions easily. 

\addcontentsline{toc}{section}{References}
\bibliography{main}
\end{document}